\documentclass[prl,twocolumn,preprintnumbers,nofootinbib]{revtex4-1}%
\pdfoutput=1

\usepackage[english]{babel}
\usepackage{epsf,epsfig}
\usepackage{amssymb,amsmath,amsfonts}
\usepackage{mathrsfs} 
\usepackage{color}
\usepackage{enumitem}

\def\m{\mathfrak{m}}
\def\nntot{n_{tot}}
\def\Sp{\mathbb{S}}
\def\Tr{{\rm {Tr}}}
\def\partition{r}
\def\hh{h}

\def\EE{{\cal E}}
\def\FF{{\cal F}}

\def\HH{{\cal H}}

\def\MM{{\cal M}}
\def\NN{{\cal N}}
\def\OO{{\cal O}}

\def\RR{{\cal R}}
\def\SS{{\cal S}}

\def\XX{{\cal X}}
\def\YY{{\cal Y}}

\newcommand{\beq}{\begin{equation}}
\newcommand{\eeq}{\end{equation}}
\newcommand{\bea}{\begin{eqnarray}}
\newcommand{\eea}{\end{eqnarray}}

\newcommand{\nn}{\nonumber}

\begin{document}

\title{Quasi-Normal Modes from Non-Commutative Matrix Dynamics}

\author{Francesco Aprile}
\email{F.Aprile@soton.ac.uk}
\affiliation{STAG Research Centre, Mathematical Sciences, School of Physics and Astronomy, University of Southampton, Highfield SO17 1BJ, UK}

\author{Francesco Sanfilippo}
\email{F.Sanfilippo@soton.ac.uk}
\affiliation{School of Physics and Astronomy, University of Southampton, Highfield SO17 1BJ, UK}

\begin{abstract}
\noindent 
We explore similarities between the process of relaxation in the BMN
matrix model and the physics of black holes in AdS/CFT.
Focusing on Dyson-fluid solutions of the matrix model, we perform
numerical simulations of the real time dynamics of the system.
By quenching the equilibrium distribution we study quasi-normal
oscillations of scalar single trace observables, we isolate the lowest
quasi-normal mode, and we determine its frequencies as function of the
energy.
Considering the BMN matrix model as a truncation of $\mathcal{N}=4$
SYM, we also compute the frequencies of the quasi-normal modes of the
dual scalar fields in the AdS$_5$-Schwarzschild background.
We compare the results, and we find a surprising similarity. 
\end{abstract}
\maketitle

%%%%%%%%%%%%%%%%%%%%%%%%%%%%%%%%%%%%%%%%%%%%%%%%%%%%%%%
\section{Introduction}
%%%%%%%%%%%%%%%%%%%%%%%%%%%%%%%%%%%%%%%%%%%%%%%%%%%%%%%

One of the most fascinating phenomena in statistical systems is
certainly the emergence of the macroscopic laws of physics.
Worth mentioning are some results obtained from the
study of random matrices: In the seminal paper~\cite{Dyson} Dyson
showed that the dynamics of the eigenvalues of a random matrix
resembles that of a ``Coulomb gas''. More recently, Blaizot and Nowak 
pointed out~\cite{Blaizot:2009ex}  that the time evolution of the one-particle density of this Coulomb gas
is related to the Burgers equation of fluid dynamics.
In this work we consider a similar problem for the BMN matrix model~\cite{Berenstein:2002jq}, 
and we study the classical time evolution 
of an analogous ``Dyson fluid'' distribution, defined 
by the ensemble of initial conditions of each independent matrix element.
We will show that qualitatively the Dyson fluid of the BMN matrix model vibrates as a black
hole in AdS. The intuition behind the emergence of this unique type 
of collective behavior comes from the AdS/CFT correspondence,
which we briefly review in the second part of this paper.

The dynamics of the classical BMN matrix model is non-commutative, and there are
two features that have to be emphasized. In the first place, the dynamical
system is chaotic~\cite{Gur-Ari:2015rcq}, therefore during the time evolution any localized
ensemble of initial conditions will spread over the allowed phase space. 
Secondly, the expectation values of the observables 
equilibrate at late times~\cite{Asplund:2011qj},
meaning that all the elements in the ensemble populate the phase
space according to a time-independent distribution. Such equilibrium
distribution is not connected perturbatively to the vacuum
configuration of the matrix model, and we generate it numerically by
molecular dynamics.
Once the system has equilibrated, we study fluctuations
close-to-equilibrium by activating a gauge invariant quench protocol which induces
specific deformations of the equilibrium distribution. The quench
protocol is novel, and allows us to perturb the system in a controlled
manner.
Ought to the expansive nature of the Hamiltonian flow, we then expect
the ensemble to quickly approach a new equilibrium.  The observables
re-equilibrate via quasi-normal oscillations which are characterized
by a complex frequency, i.e\, a damping rate and a ringing frequency.
The lowest quasi-normal mode of twist two single trace scalar operators is
isolated, and its frequency analyzed as a function of the parameter of
the ensemble.  Our results extend and consolidate into a unified
statistical framework the interesting work of~\cite{Asplund:2012tg}.

The BMN matrix model can be obtained as a classical consistent truncation of
$\mathcal{N}=4$ SYM in $4d$. Moreover,
$\mathcal{N}=4$ SYM is famously known to admit an holographic
description at strong 't Hooft coupling. Holography provides a unique
realization of the idea that, for a class of gauge theories, the large-N limit of correlation functions
is controlled by a master field configuration
\cite{Gopakumar:1994iq}. It is unique because the master field
configuration is found to be a solution of a classical gravitational
problem in Anti-de-Sitter space with prescribed boundary conditions.
From independent field theory computations, there is nowadays 
a compelling evidence that in certain supersymmetric cases holography 
indeed provides the master field configuration at strong 't Hooft
coupling~\cite{Gomis:2008qa,Buchel:2013id,Benini:2015eyy}. However,
under the generic assumptions of the AdS/CFT duality, any solution of
the gravitation problem is in correspondence with a field configuration in the
dual field theory. In the class of finite energy solutions, the most interesting 
ones are certainly black hole solutions undergoing non trivial real-time dynamics
~\cite{Bantilan:2012vu}.

In the brane engineering of $\mathcal{N}=4$ SYM, the Dyson fluid
represents a simple statistical model for a non-commutative ensemble of matrices at finite energy, in which 
the off-diagonal degrees of freedom of the branes are fluctuating. 
The phase space interpretation of the dynamics is complex, but we can look at simpler ``geometrical" objects:
these are the distribution of eigenvalues associated to gauge invariant operators,
which we will show to be non trivial, and to have a finite large-N behavior. 
This is a very different notion of geometry compared to that offered by the AdS/CFT correspondence.
Bearing in mind the impossibility of a direct comparison, 
we would like to understand, in a dynamical setting, how much these two notions of 
geometry are far from each other, 
and in particular we would like to ask how much the process of relaxation in the BMN Dyson fluid differs from 
that of a black hole in AdS. Since the late time relaxation of a black hole is
determined by the spectrum of quasi-normal modes at the equilibrium, 
we shall answer our curiosity by considering the AdS$_5$-Schwarzschild black hole as prototype.
We will find that the parametric behavior of the quasi-normal frequencies in the two systems is surprisingly similar. 
We can interpret these similarities as an indication that a proper 
path integral formulation of the full $\mathcal{N}=4$ Dyson fluid
might shed new light on the ensemble of black hole microstates.
Even though a-priori our comparison has been rather disparate,
another possibility, which we have not explored in this paper, would be to study 
the expectation value of operators of large $R$-charge, as the BMN operators, 
and look for the limiting behavior of their quasi-normal frequencies. 
We should emphasize that the computation we have performed 
in the matrix model in order to extract the quasi-normal frequencies is highly non trivial, 
and numerically challenging. 
The same would be true for the quasi-normal frequencies 
in the dual black hole background, since the latter is only known numerically \cite{Costa:2014wya}.

The rest of the paper is organized as follows: In Section~${\bf I}$,
we fix our notation and we introduce the BMN matrix model.
In Section~${\bf II}$ we describe the statistical framework which will
be used to define our Dyson fluid.  In particular, we discuss our
``big-bang'' initial condition, and we clarify some important aspects
concerning the interplay between observables, gauge invariance, and
non-commutativity.
In Section~${\bf III}$ we study the dynamics of the system from the
``big-bang'' to the equilibration: We explain how the process of
equilibration is related to chaos, and we prove that the joint
probability distribution at the equilibrium does not factorize.
In Section~${\bf IV}$ we describe our gauge invariant quench
protocol, and we then study the quasi-normal oscillations of the
scalar operators in the ${\bf 20}$ of $SO(6)$.
In Section~${\bf V}$, we take a short detour into the AdS/CFT, 
and we compute in the AdS$_5$-Schwarzschild black hole background
the quasi-normal modes of the scalar fields representing the ${\bf 20}$ of $SO(6)$.
In Section~${\bf VI}$ we compare the behavior of the
the frequencies of the quasi-normal
modes in the Dyson fluid and in the black hole background. 
Finally, in Section~${\bf VII}$ we conclude with a summary and an outlook.

%%%%%%%%%%%%%%%%%%%%%%%%%%%%%%%%%%%%%%%%%%%%%%%%%%%%%%%
\section{I.\, The BMN Matrix Model}
%%%%%%%%%%%%%%%%%%%%%%%%%%%%%%%%%%%%%%%%%%%%%%%%%%%%%%%

We consider a set of $N\times N$ hermitian matrices $X_I(t)$ and the action: 
\bea
\mathcal{S}_{BMN}&=&\int dt~\sum_I\, \mathscr{L}^{}_I - \sum_i \,
\mathscr{V}^{}_i - \sum_a\, \mathscr{V}^{}_a ~,\label{dyna_sys}
\eea
where
\bea
\mathscr{L}^{}_I&=& \Tr\,\Big[ \begin{array}{l}\frac{1}{2}\end{array}
  (D_t X_I)^2
 \begin{array}{l}+\ \frac{1}{4}\sum_J\end{array} [X_I, X_J]^2 \, \Big]\,,\label{dyna_sys_1} \\
\mathscr{V}^{}_i &=&\Tr\,\Big[\begin{array}{l}\frac{1}{2}\left(
    \frac{\m}{6} \right)^2 \end{array} X_i^2
  \Big]\,, \label{dyna_sys_2}\\
\mathscr{V}^{}_a&=&\Tr\,\Big[ \begin{array}{l}\frac{1}{2}\left(\frac{\m}{3}\right)^2 \end{array}
  X_a^2\begin{array}{l} +\,\frac{\m}{3} \sum_{bc}\end{array} i
  \varepsilon_{abc}\, X_a X_b X_c \Big]\,,\qquad \label{dyna_sys_3}
\eea
and $D_t X_I\equiv \partial_t X_I - i [A_t, X_I]$. The notation is as
follows. No distinction is made between an upper and a lower index.
We use capital letters $I,J$, to label the whole set of matrices
$\{1,\ldots,\nntot\}$ with $\nntot=9$, we use instead small letters of
the type $a$ and $i$ to label two complementary subsets of
$\{1,\ldots,\nntot\}$, respectively: $1\le a \le \partition$ with
$\partition=3$ and $\partition+1\le i \le \nntot$. The generalization
to arbitrary $\nntot$ and $\partition$ is straightforward.
The values of $\nntot=9$ and $\partition=3$ correspond to the bosonic truncation of
the BMN supersymmetric matrix model~\cite{Berenstein:2002jq}. The
Lagrangian
\beq
\mathcal{S}_{BFSS}=\int dt~ \sum_I\, \mathscr{L}^{}_I
\eeq
is known as the BFSS matrix model~\cite{Banks:1996vh}. 
Mass terms in the BMN matrix model will not be important for the dynamics of the Dyson fluid, 
the main differences between BFSS and BMN come from the cubic interaction.

The action~\eqref{dyna_sys} is invariant under local $U(N)$ gauge
transformations,
\bea
X_I &\ \rightarrow\ & U X_I U^\dagger~,\label{gi_X}\\ A_t
&\ \rightarrow\ & U A_t U^\dagger - i (\partial_t U)
U^\dagger~.\label{gi_At}
\eea
with $U(t)$ a generic matrix in $U(N)$. The BFSS matrix model has
maximal $SO(9)$ global symmetry, whereas the global symmetry group of
the BMN matrix model is reduced by the mass terms to $SO(3)\times
SO(6)$.

%==================================================================================
\subsection{Equations of Motions}
%==================================================================================

The Euler-Lagrange equations for the matrices $X_I$ are:
\bea
\label{eqxi}
\rule{0pt}{.5cm} D_t^2\, X_i &=& \sum_{J} [ X_J,[X_i, X_J]]
- \begin{array}{l} \left(\frac{\m}{6}\right)^2 \end{array} X_i~,
\rule{1cm}{0pt}\label{eqXi}\\
\label{eqxa}
D_t^2\, X_a &=& \sum_{J} [ X_J,[X_a, X_J]] - \begin{array}{l}
  \left(\frac{\m}{3}\right)^2 \end{array} X_a + \nn \rule{1cm}{0pt}
\\ & & \rule{3cm}{0pt} - i\m\, \varepsilon_{abc} X_b
X_c~. \label{eqXa}
\eea
Degrees of freedom described by $\Tr[X_I]$ are completely decoupled,
in particular
\bea
\partial_t^2\, \Tr[X_I] &=& - \begin{array}{l} \big( \sum_{j}
  \left(\frac{\m}{6}\right)^2 \delta_{jI} + \sum_{a}
  \left(\frac{\m}{3}\right)^2 \delta_{aI} \big) \end{array}\Tr
        [X_I]~. \nn
\eea
Hence, the set of interacting degrees of freedom coincides with the
set of hermitian and {\it traceless}\ matrices. It is then convenient
to parametrize the $X_I$ in terms of the generators ${\bold T}^m$ in
the fundamental of $\mathfrak{su}(N)$,
\beq
X^I = \sum_{m}^{N^2-1} x^I_m\, {\bf T}^{m}~,\qquad \forall\, I~.
\eeq
The r.h.s of the equations of motions can be written as
\bea
&& D_t^2\,x^i_m \begin{array}{l} +
  \left(\frac{\m}{6}\right)^2 \end{array} x^i_m = \sum_{J,n,p} \RR^{m
  n}_J \RR^{n p}_J x^i_p \rule{1cm}{0pt} \nn\\ &&
D_t^2\,x^a_m \begin{array}{l} +
  \left(\frac{\m}{3}\right)^2 \end{array} x^a_m = \sum_{J,n,p} \RR^{m
  n}_J \RR^{n p}_J x^a_p +\nn \rule{.8cm}{0pt} \\ &&
\rule{4.6cm}{0pt} \begin{array}{l} + \frac{\m}{2} \end{array}
\varepsilon_{abc} f^{pqm} x^b_p x^c_q~,\nn
\eea
where $\RR^{m n}_J =\sum_q f^{mqn} x_q^J$ and $[{\bf T}^m,{\bf T}^n]=i
f^{mnp} {\bf T}^p$. \rule{0pt}{.5cm}

 \rule{0pt}{.4cm} The gauge field $A_t$ has no kinetic term, and its
 equation of motion becomes a constraint:
 \beq
 \label{constraint}
 \sum_I [D_t X_I, X_I] =0~.
 \eeq

The phase space $\mathcal{P}$ associated to our dynamical system is
described by all the degrees of freedom in the variables
$X_{I=1\ldots 9}$, and $P_{I=1\ldots 9}$, where
\beq
P_I\equiv \frac{\partial \mathscr{L} }{\partial(\partial_t X_I)}= D_t
X_I~.
\eeq
A point $\mathcal{X}\in\mathcal{P}$ represents a {\it
  configuration}\ of matrices and momenta denoted as
$\mathcal{X}=\{(X_I,P_I)\}_{I=1\dots 9}$, or equivalently
$\mathcal{X}=(\{\vec{x}_m\},\{\vec{p}_m\})$ where the index $m$ runs
over the basis $\{ {\bold T}^m\}_{m=1}^{N^2-1}$ of generators, and the
vector notation stands for $\vec{x}_m=(x^1_m,\ldots,x^9_m)$.

In the Hamiltonian formalism, 
\bea
\label{Hamiltonian}
\mathcal{H}&=&\begin{array}{l} \frac{1}{2}\end{array} \sum_I
\Tr\Big(P_I^2 + i A_t [X_I,P_I] \Big) \nn\\ &
& \begin{array}{l}-\ \frac{1}{4}\end{array} \sum_{IJ}\ [X_I, X_J]^2 +
\sum_i \, \mathscr{V}^{}_i + \sum_a\, \mathscr{V}^{}_a~,
\eea
and the equations of motions~\eqref{eqXi}-\eqref{eqXa} become
equivalent to the 1st order system
\bea
\partial_t X_I &=& P_I + i [A_t,X_I]\,, \label{HamEqX}\\ \partial_t P_I
&=& +i [A_t, P_I] - \sum_{J\in \{a,i\}} \frac{ \partial\,
  \mathscr{V}_J }{\partial X_I}~, \label{HamEqP}
\eea
supplemented by the constraint \eqref{constraint}: $\sum_I
[P_I,X_I]=0$.
The integral form of the equations of motion defines the Hamiltonian
{\it flow} $\ \varphi: I\times \mathcal{P}\rightarrow \mathcal{P}$,
where $I\subset\mathbb{R}$ is a interval of time, and $\varphi$ is a
map such that for each initial point $\mathcal{X}^{(0)}$ in phase
space, the path $\gamma(t):=\varphi_t(\mathcal{X}^{(0)})$ is the
unique curve with initial condition $\gamma(0)=\mathcal{X}^{(0)}$.
The flow commutes with the action of the global symmetries of the
Hamiltonian.

%==================================================================================
\subsection{Conserved Charges}
%==================================================================================

In the BMN matrix model, the conserved quantities are the energy $E$
and the Noether charges of the $SO(3)\times SO(6)$ symmetry group.
The energy is obtained by evaluating the Hamiltonian $\mathcal{H}$
given in~\eqref{Hamiltonian}. The $SO(3)$ and $SO(6)$ charges, dubbed
$L_{c=1,2,3}$ and $J_{q=1,\ldots,15}$, respectively, are
\bea
\label{Noether_Charges} L_c = \Tr\big( X_a\,\mathbb{A}_c^{ab} P_b
\big),\quad J_q = \Tr\big( X_i\, \mathbb{Y}_q^{ij}P_j\big)~.
\eea
where the matrices $\{\mathbb{A}_c\}^{c=1,2,3}$, and
$\{\mathbb{Y}_q\}^{q=1,\ldots,15}$ generate rotations in
$\mathbb{R}^3$, and $\mathbb{R}^6$, respectively.

The canonical form of $\mathbb{A}_k$ is that of an anti-symmetric
matrix whose upper triangular part has $0$ everywhere, except for $+1$
in the position $({\bar{a}}, {\bar{c}})$, corresponding to the plane
$(X_{\bar{a}},X_{\bar{c}})$ which is being rotated.  Similarly for
$\mathbb{Y}_q$.  For example, the combination $\Tr\big( X_5 P_9 - X_9
P_5\big)$ is the charge associated to rotations in the plane
$(X_5,X_9)$.  From the equations of motion
\eqref{HamEqP}-\eqref{HamEqX}, it can be explicitly checked that
\beq
\frac{d L_c}{dt} =\frac{d J_q}{dt} =0\,,\qquad \forall\ k,\,q~.
\eeq

%%%%%%%%%%%%%%%%%%%%%%%%%%%%%%%%%%%%%%%%%%%%%%%%%%%%%%%
\section{II.\,  Statistics and Dynamics}
%%%%%%%%%%%%%%%%%%%%%%%%%%%%%%%%%%%%%%%%%%%%%%%%%%%%%%%

In the previous section we have introduced the BMN matrix model as a
dynamical system focusing on various aspects of the time evolution of
a single configuration.
In this section we describe a more general framework in which the
degrees of freedom are interpreted as the microscopic elements of a
statistical ensemble.
The statistical framework that we are advocating is an example of
a``Dyson fluid'', and constitutes our starting point for the study of
non-equilibrium dynamics in the BMN matrix model.

The definition of equilibrium that we shall use throughout the paper
is the following.  Given: a set of initial conditions, and an algebra
of observables, a dynamical system is said to be at the equilibrium
w.r.t.\ the given initial conditions, if the expectation value of any
observable in the algebra is time independent.
Let us point out that the when a statistical equilibrium is reached,
the microscopic degrees of freedom are not necessarily steady, and in
fact they can have an highly non trivial dynamics.  As we are going to
show, the process of equilibration in the BMN matrix model has
precisely this feature.

Before presenting our numerical results we describe in great detail
our choice of initial conditions, and we emphasize some important
properties of the observables that we shall study. Several aspects of
our analysis will be generic, therefore we expect our results to
provide the key elements towards the understanding of the process of
equilibration in non-commutative dynamical systems.

%==================================================================================
\subsection{Big-Bang Initial Conditions}
%==================================================================================

We consider the following class of out-of-equilibrium initial conditions:
\beq
\label{ini_cond}
A_t=0~,\qquad X_I=0~,\qquad P_I\in~ {TGU}~,
\eeq
where TGU stands for ``Traceless Gaussian Unitary ensemble''. In
practice, the momenta are parametrized as:
\beq
\label{ini_momenta_nc}
P^I = \sum_{m}^{N^2-1} p^I_m\, {\bf T}^{m}\qquad \forall\, I~,
\eeq
and each coefficient $p^I_m$ is extracted randomly from a single
gaussian distribution, which we take to have standard deviation
$\sigma$ and zero mean. Because of this property, we expect
$[P_I,P_J]\neq 0$ for all of the randomly generated initial
conditions, apart from a set of zero measure. The
constraint~\eqref{constraint} is satisfied at the initial time, and
therefore $A_t$ will remain zero during the time evolution.

Intuitively, we designed the initial conditions~\eqref{ini_cond}
having in mind a {\it big-bang} in which the variables $\{X_I\}$ start
at the origin with random momenta. As soon as the $\{X_I\}$ do not
commute, interactions are turned on. This can be understood by
looking at the force terms in~\eqref{eqXi}-\eqref{eqXa}, namely
\bea
\FF^{(3)}_a &\propto& i\varepsilon_{abc} X_b X_c~, \nn\\
\FF^{(4)}_I &\propto& \begin{array}{l} \sum_{J}\end{array} [ X_J,[X_I,
    X_J]]~.
\eea
For a given initial condition we shall find $[P_I,P_J]\neq 0$ at the
initial time, and after an infinitesimal time step of $\delta t$,
\beq
X_I (\delta t) = X_I (0) + \delta t\,P_I(0)=\delta t\,P_I(0)~,
\eeq
thus $\FF^{(3)}_a\neq\FF^{(4)}_I\neq0$. With these forces turned on,
the degrees of freedom are coupled, and will evolve non trivially
under the Hamiltonian flow $\varphi$.
 
We generate a finite set of initial conditions of the
form~\eqref{ini_cond}-\eqref{ini_momenta_nc}, which will be denoted by
$\mathcal{E}(0)$, where $\mathcal{E}$ stands for {\it ensemble}. By
definition $\mathcal{E}(t)=\varphi_t(\mathcal{E}(0))$.  Notice that
$\mathcal{E}(0)$ belongs to a Lagrangian submanifold of the phase
space, centered at $X_I=P_I=0$, and extended only along the directions
of the momenta.
The numerical integration is achieved through an improved second order
leap-frog algorithm~\cite{Omelyan}. The time step used in the
simulation is $\delta t= 0.05$. The stability of the integration has
been checked by increasing and decreasing $\delta t$ of a factor of
two.
Once the ensemble $\mathcal{E}(t)$ is obtained, we compute the
expectation value of any observable $\mathcal{O}$ through the
estimator,
\beq\label{def_OOt_num}
\langle \OO \rangle(t) \equiv \frac{1}{vol(\mathcal{E}(t))}
\sum_{\mathcal{X}\in\,\mathcal{E}} O(\mathcal{X})~,
\eeq
where $vol(\mathcal{E}(t))$ equals the number of configurations, and
we determine the error on $\langle \OO \rangle$ by means of a
jackknife analysis.

On $\mathcal{E}(t)$, the $SO(3)\times SO(6)$ charges vanish because
the $\{X_I=0\}$, whereas the energy is non zero. Since the energy is
purely kinematical at the initial time, we can calculate its
expectation value and its standard deviation, analytically,
\beq\label{Einitial}
\begin{array}{ccl}
\langle E \rangle &=&~\nntot (N^2-1)~ \sigma^2~,
\\ \rule{0pt}{.5cm}SD[E]&=& \sqrt{2\, \nntot (N^2-1) } ~ \sigma^2~.
\end{array}
\eeq
It is convenient to define the parameter $h$ and rescale $\sigma$ in
such a way that $\langle E\, \rangle$ does not depend on $N$,
\beq
\sigma= \sqrt{\frac{\hh}{N^2-1}}~.
\eeq
Then, $\langle E\, \rangle=\hh\, \nntot $.  For concreteness we will
always specify the dependence on the expectation value
of the energy by referring to $\EE=\EE_h$.

The statistical framework developed so far has a general validity.
For the BFSS matrix model, the study of the ensemble greatly
simplifies because when $\mathfrak{m}=0$ the
equations of motion enjoy the scaling symmetry: $t\rightarrow
\lambda^{-1} t$ with $X_I\rightarrow \lambda
X_I$~\cite{Asplund:2012tg}.  In particular, if two configuration in
phase space are related as $\mathcal{X}_2=\lambda \mathcal{X}_1$, we
find that
\beq\label{Symm_BFSS}
\varphi_{t/\lambda} (\mathcal{X}_2)=\varphi_t(\mathcal{X}_1)\,,\qquad
E[\mathcal{X}_2]= \lambda^4 E[\mathcal{X}_1]~.
\eeq
Therefore, given one ensemble of fixed $h$, any other ensemble 
of finite energy can be generated by means of the scaling symmetry.  
The BMN matrix model instead, depends on the value of $\m$ 
in a non trivial way.  The strategy in this case is to keep fixed 
the value of $\m$, and generate the one-parameter family
$\mathcal{E}_h$ by varying $h$. There is no loss of generality in
doing so, because ensembles in which $\m$ has a different value can be
obtained by the rescaling:
$t\rightarrow \lambda^{-1} t$, $X_I\rightarrow \lambda X_I$ with
$\m\rightarrow \lambda \m$.
Let us mention that $\sigma\rightarrow \lambda^2\sigma$ under the
scaling, because $\sigma$ generates momenta.  In the numerical
simulations, we will fix the value of the mass to $\m=3$.

%=================================================================================================
\subsection{Commuting Solutions}
%=================================================================================================

Non interacting solutions are described by the commuting ansatz
$[X_I,X_J]=0$, $\forall$ $I$, $J$.  In this case the equations of
motion admit a simple set of time-periodic solutions.  It is useful to
identify such solutions, since comparing commuting versus non-commuting
solutions will also clarify in which sense the latter are different.
 
Assuming $[X_I,X_J]=0$, a common basis of eigenvectors exists such
that the $X_{I=1,\ldots9}$ are diagonal.  The diagonal degrees of
freedom are decoupled and become simple harmonic oscillators.
Assuming $X_I=0$ at the initial time, $t=0$, we find
\bea\label{comm_sol}
(X^I)_{mn} &=& x^I_m(t)\, \delta_{mn}~, \\ x_m^a &=& \begin{array}{l}
  \frac{3}{\m} \end{array} p^a_m \begin{array}{l} \sin
  \left(\frac{\m}{3}\, t\right)\,,\end{array}~ x_m^i = \begin{array}{l}
  \frac{6}{\m} \end{array} p^i_m \begin{array}{l} \sin
  \left(\frac{\m}{6}\, t\right)\,,\end{array} \nn \rule{0pt}{.5cm}
\eea
where $p^I_m$ are the initial velocities, constrained by the
requirement that $X^I$ is traceless. In the limit $\m\rightarrow 0$
we recover solutions of the BFSS matrix model with the given initial
conditions,
\beq
(X^I)_{mn} =  x^I_m(t),~\quad
x_m^I=  p^I_m~t~.
\eeq
In the BFSS model, constant commuting matrices $X_I$ parametrize the
flat directions of $\FF^{(4)}$ (and $\FF^{(3)}$). Such flat
directions are lifted by the mass terms in the BMN potential, and
therefore the solution acquires a non trivial time dependence.
Let us mention that the BMN potential admits a set of zero energy
vacua known as {\it fuzzy spheres}. These vacua are labelled by
adjoint $\mathfrak{su}(2)$ representations, and they are defined by
the conditions $X_i=0$, $[X_a,X_b]=i\varepsilon_{abc} X_c$. The
Hamiltonian flow of fuzzy spheres configurations has been discussed
in~\cite{Asplund:2011qj}.

%==================================================================================
\subsection{Local Observables}
%==================================================================================

The dynamics of $\mathcal{E}_h(t)$ will be monitored by measuring the
expectation value of gauge invariant observables as function of time.
The simplest observables are the kinetic energy,
\beq
K=\begin{array}{l} \sum_I \Tr(P_I^2)\,,\end{array}
\eeq
and the singlets of $SO(3)$ and $SO(6)$, 
\bea
O_s^{(3)}\equiv \sum_a \Tr(X_a^2)\,,\quad
O_s^{(6)}\equiv \sum_i \Tr(X_i^2)~.
\eea
More interesting operators belong to non trivial irreducible
representations (irrep) of the global symmetry group.  In this work,
we will consider mainly the symmetric traceless irrep of $SO(6)$,
\beq
\OO_{ij} = \Tr[ X_i X_j ] \begin{array}{l} -\frac{1}{6} \sum_k
  \Tr[X_k^2]\end{array} \delta_{ij}~,
\eeq
of dimension ${\bf 20}$. This irrep can be conveniently decomposed
according to the $SU(3)$ subgroup of $SO(6)$, as ${\bf 8}\oplus{\bf
  6}\oplus {\bf \bar{6}}$.  In particular, we will focus on,
\bea
\NN_{1} &\equiv& \Tr ( X_4^2+X_5^2 - X_6^2-X_7^2)\,,\\ \NN_{2}
&\equiv& \Tr ( \begin{array}{l} \sum_{i=4}^7 X_i^2\end{array} -2
  X_8^2-2X_9^2)\,,\qquad\\ \mathcal{C}[X_i,X_{i+1}] &=&\Tr \big[\,
    (X_i+i X_{i+1})^2\big],\quad i=4,6,8\,.
\eea
The $\NN_{i=1,2}$ are singlets under the maximal torus $U(1)^3\subset
SO(6)$ and belong to the (real) irrep of dimension ${\bf{8}}$ of
$SU(3)$. The $\mathcal{C}_{i=4,6,8}$ are instead charged under the
$U(1)^3$, and belong to the ${\bf 6}\oplus {\bf \bar{6}}$ of $SU(3)$.
Finally, the conserved charges $\{ L_c\} $ and $\{J_q\}$ are
(time-independent) observables transforming in the adjoint
representation of $SO(3)$ and $SO(6)$, respectively.

In general, gauge invariant observables are of type
\beq\label{def_obs}
\OO = \Tr(\,\MM[ \{X_I\}]\,)~,
\eeq
where $\MM$ is polynomial which may depend both on the coordinates
$\{X_I\}$ and the momenta $\{D_tX_I\}$. Gauge invariance is ensured
by the transformation law
\beq
\MM[\, U\{X_I\}U^\dagger] = U\MM[ \{X_I\}] U^\dagger~.
\eeq
In this formalism, the operators $\OO_s$ and $\NN_{i=1,2}$ have a
corresponding $\MM$ of the form
\beq\label{ex_herm_obs}
\sum_{I} c_I f_I(X_I)\,,\qquad c_I\in\mathbb{R}~,
\eeq
with $f_I$ a simple polynomial.  When an observable $\OO$ is defined
as in~\eqref{def_obs}, and furthermore the matrix $\MM$ is hermitian,
calculating $\OO$ is equivalent to summing over the real eigenvalues
of $\MM$.  Two different observables, defined by $\MM_1$ and $\MM_2$,
respectively, will not in general have a common basis of eigenvectors,
because the $\{X_I\}$ do not commute.  However, even though the
eigenvectors of $\MM_{i=1,2}$ have no intrinsic meaning, the
eigenvalues can be understood as particular functions of the dynamical
degrees of freedom $\{x^I_m\}$, and as such, studied on their own. In
this case, an interesting quantity to consider is the distribution of
eigenvalues of a given $\MM$, integrated over the ensemble.

The relation between the eigenvalues of the $\{X_I\}$, the
observables, and the gauge invariance of the system, is more delicate.
Some remarks are in order:\\

\begin{enumerate}[wide, labelwidth=!, labelindent=0pt]
  \item Given a configuration $\XX$, we may choose to diagonalize one (and
    only one) of the matrices $\{X_I\}$ by acting with a local gauge
    transformation $U(t)$. The eigenvalues of this matrix are promoted
    to dynamical variables, and the price we pay is to introduce a non
    trivial gauge field, $A_t=-i (\partial_t U) U^\dagger$. The
    rotated configuration, $\XX^{(U)}(t)$, carries the same
    information of $\XX(t)$. \\
  
  \item As we mentioned, observables whose matrix $\MM$ is hermitian,
    are sensitive only to the eigenvalues of $\MM$.  A stronger
    statements holds for observables whose matrix $\MM$ has the form
    given in~\eqref{ex_herm_obs}. In this case, the observables are
    sensible only to the eigenvalues of the $\{X_I\}$, even though
    dynamically it is not possible to diagonalize simultaneously all
    the matrices.  We could say that the eigenvalues of the matrices
    $\{X_I\}$ represents the first coarse-grained variables built
    out of the microscopic degrees of freedom $\{x^I_m\}$.
\end{enumerate}

%================================================================================
\section{III. Aspects of Equilibration}
%================================================================================

\begin{figure}[t]
\includegraphics[scale=0.33]{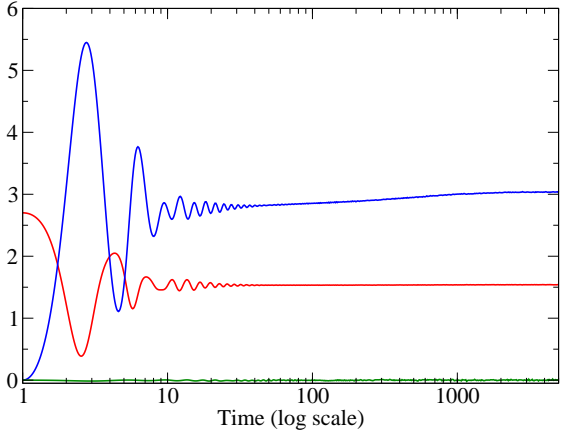}
\vskip 0cm
\caption{From top (red) to bottom (blue), expectation values of
  $\OO_h^{(6)}$, $K$, and $\NN_1$, for ${h=0.3}$ with $N=15$ and
  $vol(\EE)\approx 10^3$.  In the log coordinate $t=\log s$, the approach to
  the equilibrium is $\langle\OO\rangle(t) - \OO_{\infty} = d/s^\tau$
  with $d>0$ for $\OO_h^{(6)}$ and $d<0$ for $K$.}
\label{Fig1}
\end{figure}

From the numerical results it is clear that the big bang is a
far-from-equilibrium initial condition, but also that the system
equilibrates at late times.
The typical behavior of the observables is exemplified in
Figure~\ref{Fig1}: Right after the big-bang these observables
experience a highly non linear dynamics characterized by a sequence of
oscillations, rapidly enough these oscillations relax and their
expectation values settle down.

At the initial time $\langle \OO_s^{(6)}\rangle=0$, and $\langle K
\rangle$ coincides with~\eqref{Einitial}. The late time behavior of
these two observables fits into the ansatz,
\beq\label{estimate_tau}
\langle\OO\rangle(t) = \OO_{\infty} + d\, e^{- t/\tau }, \quad \OO=K,\,\OO_s^{(6)}\,,
\eeq
with $\tau\approx 700$ in units of time. 
The dependence of $\tau$ as function of $h$ can be inferred by
dimensional analysis. The scaling we should keep in mind is
\beq\label{scalingBMN}
t\rightarrow \lambda^{-1}t\,,\qquad \{\mathfrak{m}, X_I\} \rightarrow
\lambda \{\mathfrak{m}, X_I\}~.
\eeq
In the case $\mathfrak{m}=0$, we deduce immediately that $\tau h^{1/4}
= C(N)$, where $C$ is a dimensionless constant which only depends on
$N$.  When the mass is non vanishing, another dimensionless ratio can
be considered.  In this case the right parametrization is
\beq\label{formula_tau}
\tau h^{1/4} = \Big[ C(N) + f\left( \frac{\mathfrak{m}}{h^{\,1/4}}
  ,N\right) \Big]~,
\eeq
with $f$ a function such that $f\rightarrow 0$ as $h^{1/4}\gg
\mathfrak{m}$. The same scaling argument can be used to infer the
expectation value at the equilibrium of any observable $\OO$ of
dimension $\delta$, as function of $h$ and $\mathfrak{m}$:
\beq\label{O_geneal_h_m}
\langle \OO\rangle = h^{\delta/4}\Big[ C(N) +
  f\left(\frac{\mathfrak{m}}{h^{\,1/4}} ,N\right) \Big]~.
\eeq
Both these formulae contain implicitly the assumption that the system
equilibrates.  We observed experimentally that the regimes of
applicability are two:
\begin{enumerate}[wide, labelwidth=!, labelindent=0pt]
\item the {\it asymptotic} regime $h^{1/4}\gg\mathfrak{m}$;
\item the intermediate regime $h^{1/4}\sim\mathfrak{m}$.
\end{enumerate}
In the first case the dynamics tends to the BFSS matrix model because
as $h\rightarrow \infty$ the mass term in the BMN Lagrangian is
negligible. In the second case instead, corrections due to the mass
term become important. The case $h^{1/4}\ll\mathfrak{m}$ is more
complicated, since the mass term
are not suppressed, and the time evolution could be comparable 
to that of harmonic oscillator (possibly for a time longer than our simulation time). 
In the next sections we will consider numerical simulations 
in which either 1) or 2) are valid.

The time evolution of the $\langle \OO_s^{(3)}\rangle$ singlet is similar to that of $\langle \OO_s^{(6)}\rangle$.  
The trivial result $\langle\NN_{1} \rangle=0$, also shown in Figure~\ref{Fig1}, was expected.
Indeed, the time evolution commutes with the action of the
global symmetries, and since the initial data at the big-bang are
determined by a single gaussian, $\mathcal{E}_h$ is symmetric under
$SO(3)\times SO(6)$.  It follows that for any configuration
$\mathcal{X}\in \mathcal{E}_h$, and group element $g$, the `rotated'
configuration $g\cdot\mathcal{X}$ also belongs to $\mathcal{E}_h$,
therefore, the expectation value of any charged observable vanishes.
The results obtained so far extend in different directions those
obtained in~\cite{Asplund:2011qj,Asplund:2012tg}.

%================================================================================
\subsection{Equilibration and Chaos}
%================================================================================

Some initial conditions do not lead the system to a late time
equilibrium. For example, the expectation value of any observable
evaluated on the solutions~\eqref{comm_sol} is obviously periodic in
time. Thus, the process of equilibration must be triggered by the
non-commutative interaction in the Lagrangian. However, this
microscopic feature of the system cannot by itself explain the
statistical equilibration of the expectation values of the
observables, which is instead a collective phenomena.  As we are going
to see, {\it chaos} is the key towards our understanding of the
process of equilibration.

In order to relate chaos and equilibration, we have to discuss
an important point.  On one hand, the BMN Hamiltonian is non
dissipative, and any configuration in the ensemble will keep evolving
under time evolution.  On the other hand, the expectation value of the
observables becomes time independent at late times.  It is useful to
resolve this apparent logical difficulty by first considering a
simpler situation: We may decide to calculate
\beq\label{ex_thermo_allps}
\langle\mathcal{O}\rangle_0 = \frac{1}{vol(S)}\int_S
d\mu\ \mathcal{O}(\{\vec{p}_m\},\{\vec{x}_m\})~,
\eeq
where $S$ is the submanifold in phase space allowed by the conserved
charges, and $d\mu$ is the flow-preserving volume form on $S$.  Acting
with $\varphi_t$ on $S$ we may calculate as well
\beq
\langle\mathcal{O}\rangle_t=\frac{1}{vol(\varphi_t(S))}\int_{\varphi_t(S)}
d\mu\ \mathcal{O}(\{\vec{p}_m\},\{\vec{x}_m\})~.
\eeq
Even though the flow will move any single point in $S$, because
$\varphi_t(S)=S$, we conclude that
$\langle\mathcal{O}\rangle_t=\langle\mathcal{O}\rangle_0$, and
therefore $\langle\mathcal{O}\rangle$ is at the equilibrium.  The same
conclusion would still hold true, if instead of $S$, we consider a
region $\mathcal{A}$ dense in $S$, and we assume the flow to be such
that for any $t\ge t_{eq}$ the set $\varphi_t(\mathcal{A})$ is
uniformly distributed in $S$.  In the most general situation,
equilibration takes place under the weaker condition that
$\varphi_t(\mathcal{A})$ at late times defines a probability
distribution which is time independent.  This distribution, called
$\mathscr{P}_t$ hereafter, is obtained in the statistical sense by
taking the continuum limit in the volume of $\mathcal{E}$ at fixed
time, i.e.
\bea
&& \frac{1}{vol(\mathcal{E}(t))} \sum_{\mathcal{X}\in\,\mathcal{E}}
\OO(\mathcal{X})\rightarrow \nn\\ &&\rule{1cm}{0pt} \int_S d\mu\,
\mathscr{P}_t(\{\vec{p}_m\},\{\vec{x}_m\})\ \OO(\{\vec{p}_m\},\{\vec{x}_m\})\,.
\eea
It is therefore well defined both for small and large-N.  The case of
an uniform $\mathscr{P}_t$ at the equilibrium is the case we mentioned
in the example given above.

In our finite $N$ simulations, we should also bear in mind another
detail: The fluid, $\EE_h(t)$, is not restricted to a single
submanifold $S$ but covers the set $\cup_{\gamma} S_{\HH}$ where $\HH$
is the energy of a single configuration in the ensemble.  In fact, the
energy of the ensemble is a gaussian variable with mean $\langle
E\rangle= h\, n_{tot}$ and standard deviation $\sqrt{2}\, \langle
E\rangle/ \sqrt{n_{tot} (N^2-1)}$, as shown in~\eqref{Einitial}.

The property we have invoked about the flow is very much related to
the definition of Lyapunov chaos~\cite{NoteLyapu}. An Hamiltonian
flow $\varphi$ is said to be chaotic in the sense of Lyapunov if two
properties hold:
\begin{enumerate}[wide, labelwidth=!, labelindent=0pt]
\item
  $\varphi_t$ is {\it almost everywhere expansive},
\item
  under time evolution $\varphi_t(p)$ visits almost every point in
  $S$, i.e. $\varphi_t$ is topological transitivity\footnote{A flow $\varphi$ is said to be expansive if there
  is a constant $\delta>0$ such that for any pair of points $x\neq y$,
  in $S$, there exists a time $t$ for which
  $d(\varphi_t(x),\varphi_t(y)) > \delta$.  A flow is topologically
  transitive if there exists a point $x\in S$ such that its orbit is
  dense in $S$.  By the Birkhoff transitivity theorem a flow $\varphi$
  is topologically transitive iff for any two open sets $U$ and $V$ in
  $S$, there exists a time $t$ such that $\varphi_t(U)\cap V\neq0$.}.
\end{enumerate}
In particular, the well known Lyapunov exponent $\lambda_{Lya}$
quantifies how much $\varphi_t$ is expansive. We have measured
$\lambda_{Lya}$ in our simulations, by using the strategy
outlined in~\cite{Gur-Ari:2015rcq}, for the BFSS matrix model.  Within
error we obtain the same result: $\lambda_{Lya}\approx
0.29+ O(1/N)$.
Even though the ensemble $\mathcal{E}(0)$ was gaussian and localized
in phase space, as a consequence of the expansive nature of the flow,
$\varphi_t(\mathcal{E}(0))$ populates the allowed phase at late times.
This is an important feature of the dynamics of chaotic systems as
opposed to that of integrable systems. In fact, by taking a constant
distribution of initial conditions, it is not difficult to build a
fine-tuned equilibration even for the harmonic oscillator.  However,
this type of equilibration is non generic and will fail upon
specifying a different set of initial conditions.  In this sense,
equilibration for chaotic systems is generic for any given initial
conditions.

%==================================================================================
\subsection{Dynamics of Eigenvalues}
%==================================================================================

Building on our previous remarks about the dynamics of observables of
the type
\beq\label{geo_obs}
\OO = \sum_{I} c_I\, \Tr (\, f_I(X_I) )\,,\qquad c_I\in\mathbb{R}~,
\eeq
we can re-express the expectation value of these observables as the
integral
\beq\label{prescr1}
\langle \OO \rangle = \int \prod_{I=1}^N d\lambda^J_{\alpha}
\ \mathscr{P}^{eig}(\{\lambda^J_{\alpha}\} ) \, \sum_{I,\alpha}
c_If_I(\lambda^I_{\alpha})~,
\eeq
where $\alpha=1,\ldots N,$ and
$\mathscr{P}^{eig}(\{\lambda_\alpha^J\})$ is the joint probability
distribution of all the eigenvalues of the matrices $\{X_I\}$.
The summation over the indexes $I$ and $\alpha$ can be brought outside
the sign of integration, and the expectation value in~\eqref{prescr1}
is then determined only from the knowledge of the {\it mean eigenvalue
  density} $\rho_I^{eig}(\lambda)$ of the $I$-th matrix.  This density
is obtained by integrating both the set of $\{\lambda^{\alpha}_J\}$
with $I\neq J$ and the set of $N-1$ eigenvalues in
$\{\lambda^{\alpha}_I\}$.
Assuming the existence of an equilibrium, the $SO(3)$ and $SO(6)$
symmetries of the ensemble $\mathcal{E}(t)$ imply the relations
\beq
\rho^{eig}_i(\lambda)=\rho^{eig}_j(\lambda)\,,\qquad
\rho^{eig}_a(\lambda)=\rho^{eig}_b(\lambda)\,,
\eeq
$\forall\ i,j,a,b,$ so that only two non trivial eigenvalue
distributions exists, one for each symmetry group.  Accordingly, the
charged operators $\mathcal{N}_{i=1,2}$ have vanishing expectation
values.  In the BFSS model the symmetry group is enhanced to $SO(9)$
and we would find in addition that ${\rho}^{eig}_i={\rho}^{eig}_a\ $
$\forall\ i,a$.  In the BMN model $\mathfrak{m}\neq0$, and we do not
expect this relation to hold. In fact, in the setup of
Figure~\ref{Fig1}, we can explicitly verify that
$\langle\OO_s^{(6)}\rangle\neq 2 \langle\OO_s^{(3)}\rangle$.

\begin{figure}[t]
\includegraphics[scale=0.38]{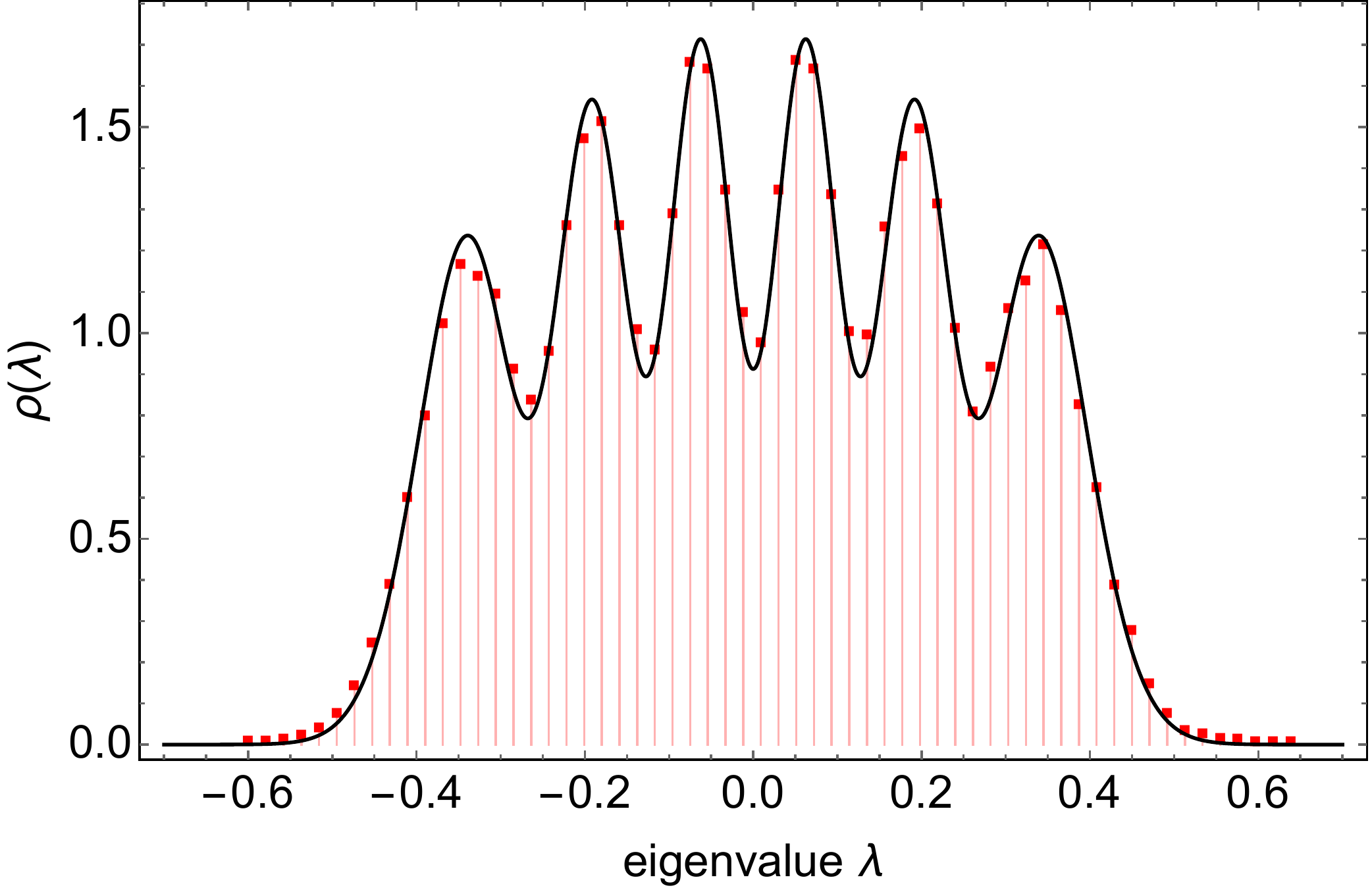}
\vskip 0.cm
\caption{The histogram of the distributions $\rho^{eig}$ in the
  $SO(6)$ sector for $N=6$ (top) and $h=0.3$. The solid black line is
  the matched TGU distribution. $vol(\EE)\approx 10^5$. }
\label{FigExtra1}
\end{figure}

\begin{figure}[t]
\includegraphics[scale=0.32]{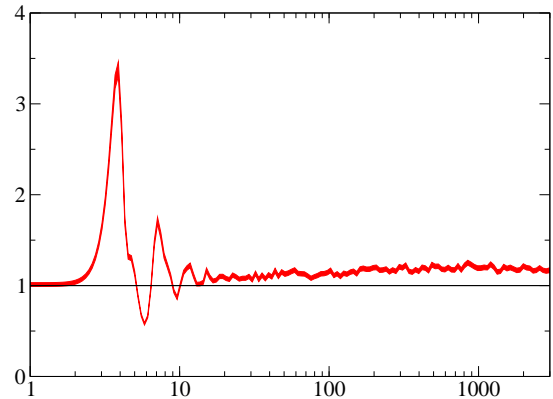}
\vskip -.8cm
\includegraphics[scale=0.32]{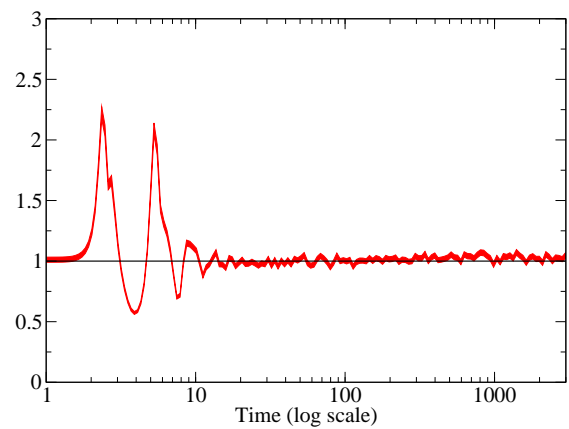}
\vskip -0.5cm
\caption{Time evolution of $\YY[X_3,X_4]$ (top) and $\YY[P_3,P_4]$
  (bottom) for $N=10$ and $h=0.3$, with $vol(\EE)\approx 10^4$.  The
  solid line $\YY=1$ is the TGU result.}
\label{FigExtra2}
\end{figure}

The histogram of $\rho^{eig}(\lambda)$ in the $SO(6)$ sector for $N=6$
and $\mathcal{E}_{h=0.3}$ is shown in Figure~\ref{FigExtra1}.  The
solid curve is the TGU distribution, which can be computed
analitically from the results of~\cite{TGUexact}.  The numerical and
the TGU distributions agree.  For completeness, let us illustrate the
analytic calculation: We quote from~\cite{TGUexact} the profile
function
\bea
p_6(\lambda)&=& e^{-\frac{6}{5} \lambda^2 } \big( \begin{array}{l}
  \frac{13436928}{244140625} \lambda^{10} - \frac{4478976}{9765625}
  \lambda^8 + \end{array} \\ &
&\rule{0pt}{.5cm}\rule{.1cm}{0pt} \begin{array}{l} +\frac{
    2581632}{1953125} \lambda^6 -\frac{102816}{78125} \lambda^4 +
  \frac{7812}{15625} \lambda^2 +\frac{644}{15625}\end{array}\big)~.\nn
\eea
The distribution $\rho^{eig}_{TGU}(\lambda)$ is proportional to $p_6
(r \lambda )$, where the parameter $r$ is obtained from the relation
\beq
\Tr(X^2)=N \int d\lambda\, \lambda^2 \rho^{eig}_{TGU}(\lambda)~.
\eeq
The good agreement between $\rho^{eig}_{TGU}(\lambda)$ and the
distribution of eigenvalues of any $X_I$ at the equilibrium should not
be confused with the fact that the ensemble at the equilibrium is
TGU.  In fact, as a consequence of the non trivial dynamics,
the matrices are correlated, and the joint probability distribution
does not factorize,
i.e. $\mathcal{P}(\vec{x}_m,\vec{y}_n)\neq\mathcal{P}(\vec{x}_m)\mathcal{P}(\vec{y}_n)$.
A simple convincing argument is the following: Let us assume that
$Z_1$ and $Z_2$ are uncorrelated and TGU with variance $\sigma$. Then
\bea
\langle \Tr(Z_i^2) \rangle_{TGU} & = &\, {(N^2-1)}
(2\sigma^2)\,,\\ \langle \mathcal{C}_4[Z_1,Z_2]\, \overline{
  \mathcal{C}}_4[Z_1,Z_2] \rangle_{TGU}&=& 8 (N^2-1) (4\sigma^4)\,,
\eea
and we obtain the {\it exact} relation, $\YY[Z_1,Z_2]_{TGU}=1$, where,
\beq\label{nonTGUrelation}
 \YY[Z_1,Z_2]\equiv\frac{N^2-1}{8}\frac{\langle \mathcal{C}_4[Z_1,Z_2]
   \overline{\mathcal{C}}_4[Z_1,Z_2], \rangle}{\langle \Tr(Z_1^2)
   \rangle \langle\Tr(Z_2^2) \rangle }~.
\eeq
The presence of dynamical correlations in $\EE_h(t)$ at the
equilibrium can be detected by studying, for example, the time
evolution of $\YY[X_3,X_4]$. In the top panel of
Figure~\ref{FigExtra2} we have plotted the expectation value of this
observable for $\mathcal{E}_{h=0.3}$ and $N=10$.  At the equilibrium,
$\YY[X_3,X_4]$ deviates considerably from unity, and we conclude that
$\mathcal{E}(t)$ is not TGU. Since the approach to the equilibrium is
exponentially fast, the result is robust. From dimensional analysis
we know that the value of $\YY[X_3,X_4]$ at the equilibrium (hereafter
$\YY_{eq}$) is a function of $N$ and $\mathfrak{m}/h^{1/4}$. In
particular, as $h$ is increased, we expect to recover the result at
$\m=0$ (which is $h$ independent).
From our simulations we are able to confirm the correctness of this
argument.  In fact, starting from the result of
Figure~\ref{FigExtra2}, where $\YY_{eq}\approx 1.19$ for $h=0.3$, we
have tested the converge of $\YY_{eq}$ measuring $\YY_{eq}\approx
1.26$ for $h=38.4$, and $\YY_{eq}\approx 1.31$ at $\m=0$.

Differently from the $X_I$, we can shown that the momenta belong to a
TGU ensemble.  This can be inferred from the bottom panel of
Figure~\ref{FigExtra2}, where we have measured $\YY[P_3,P_4]$.  At
late times, $\YY[P_3,P_4]-1$ is zero within the error.  Notice that at
very early times instead, the time evolution of the coordinates is
driven by that of the the momenta, and since the latter are initiated
as TGU, the value of $\YY[X_3,X_4]$ is still close to unity as long as
the non-commutative interactions can be neglected.

%%%%%%%%%%%%%%%%%%%%%%%%%%%%%%%%%%%%%%%%%%%%%%%%%%%%%%%
\section{IV.\,  Dynamical Relaxation }
%%%%%%%%%%%%%%%%%%%%%%%%%%%%%%%%%%%%%%%%%%%%%%%%%%%%%%%

The aim of this section is to study the close-to-the equilibrium
relaxation of the observables.  In order to do so, we devise a quench
protocol that takes $\mathcal{E}_h(t)$ at the equilibrium, and push it
onto a phase space region whose distance from $\mathcal{E}_h(t)$ can
be tuned in a controlled way.
We define the notion of distance between two ensembles by measuring
the differences between the conserved charges.
By assuming that the level sets of the conserved charges change in a
smooth way, a given quench will generate a close-to-equilibrium
configuration when the variation of the conserved charges before and
after the quench is small. Instead, the quench will generate a
far-from-equilibrium configuration when the conserved charges are
changed abruptly.  As we are going to show, our results will indeed
confirm this phase space picture.

%==================================================================================
\subsection{Quench Protocol}
%==================================================================================

Let us begin by describing our three-steps protocol:
\begin{enumerate}[wide, labelwidth=!, labelindent=0pt]
\item From the initial big-bang we wait until the system equilibrates, say at time $t_{quench}$.\\ 
\item We perform a global gauge transformation with an unitary matrix
  $U$ on (the history of) any configuration $\mathcal{X}$, i.e.
\bea
X_I(t)&\rightarrow& U X_I(t) U^\dagger\equiv X_I^{(U)}\,,\\
P_I(t)&\rightarrow& U P_I(t) U^\dagger\equiv P^{(U)}_I\,,
\eea
and we choose $U$ such that at $t_{quench}$, and for a given index
$\mathfrak{p}$, the matrix $UX_{\mathfrak{p}}U^\dagger=D$ is
diagonal;

\item For any $D$, we order the eigenvalues of $D={\rm
  diag}(\lambda_1,\ldots,\lambda_N)$ from least to greatest, and we
  perform the shift
\beq\label{shiftD}
D~\rightarrow~ \widetilde{D} \equiv D+
\mathrm{diag}(\epsilon_1,\ldots,\epsilon_N)~,
\eeq
with {\it quench parameters} $\epsilon_{\alpha=1,\ldots N}$.
\end{enumerate}
The constraint is not violated if at the same time we perform a
deformation of the momentum
\beq\label{shiftP1}
P^{(U)}_{\mathfrak{p}}~\rightarrow ~
\widetilde{P}^{(U)}_{\mathfrak{p}}\equiv P^{(U)}_{\mathfrak{p}} +
\ \delta[ P^{(U)}_{\mathfrak{p}},D ]~,
\eeq
where the operator $\delta[P,D]$ returns the hermitian matrix which
solves the matching condition,
\beq\label{eq_quenchP}
\sum_I [P^{(U)}_I,X^{(U)}_I] = \sum_{I\ne\mathfrak{p}}
    [P^{(U)}_I,X^{(U)}_I] + [ \widetilde{P}^{(U)}_{\mathfrak{p}} ,
      \widetilde{D}\,]~.
\eeq
The traceless condition of $\widetilde{D}$ is achieved by considering
quench parameters which add up to zero.  The traceless condition is
automatic in $\delta[P,D]$ because its diagonal elements do not play a
role in solving~\eqref{eq_quenchP}, and therefore can be chosen
arbitrarily.  The complete solution of $\delta[P,D]$ is not
illuminating, instead we shall mention two interesting cases which we use in
our numerical simulations:
\begin{itemize}[wide, labelwidth=!, labelindent=0pt]
\item Consider perturbing the edge of $D$ by taking
  $\epsilon_1=-\epsilon=-\epsilon_N$ and $\epsilon_{\alpha}=0$ for
  $\alpha\neq \{1,N\}$. The corresponding $\delta[P,D]$ returns an
  hermitian matrix whose entries are all zero but,
\bea
\nn
(
\delta[P,D])_{1m} \,= &- \frac{  \epsilon~ (P)_{1m} }{ \epsilon - (D)_{11} + (D)_{mm} }&\quad 2\leq  m< N \\
(\delta[P,D])_{1N}\,= & - \frac{  2\epsilon~ (P)_{1N} }{ 2\epsilon - (D)_{11} + (D)_{NN} } & \label{shiftP2}\\
(\delta[P,D])_{mN} =& - \frac{  \epsilon~ (P)_{mM} }{ \epsilon - (D)_{mm} + (D)_{NN} }&\quad 2\leq  m< N~.\nn 
\eea
Let us notice that in order for this deformation to actually take
place as $N\rightarrow \infty$, it may be necessary to consider,
\beq\label{DtildelargeN}
\widetilde{D} \equiv D+ \mathrm{diag}(
\underbrace{-\epsilon_1,..,-\epsilon_k},0,..,0,\underbrace{+\epsilon_k,..,+\epsilon_1})\,,
\eeq
while keeping $k/N$ fixed.  It is worth emphasizing at this point,
that the quench is not symmetric under $\epsilon\rightarrow-\epsilon$.
This can be understood by considering the denominators of $\delta[ P,D
]$. For example, the quantity $-(D_{11}-\epsilon)+D_{22}$ depends on
both the sign of $\epsilon$, and the difference between two
near-neighbour eigenvalues of $\widetilde{D}$. For one sign
$\epsilon$, there exists a small value of $-\epsilon$ for which the
denominator diverges, and as a result, we expect the quenched
configuration to be a far-from-equilibrium configuration.
\item Consider a quench in
  which $D$ is inflated. This configuration is achieved by defining
  $\epsilon_\alpha=\lambda_\alpha \overline{\epsilon}$, with a given
  $\overline{\epsilon}$. In matrix notation
\beq
\widetilde{D}=(1+\overline{\epsilon})D~.
\eeq
Notice that the traceless condition is automatically satisfied. The
solution of $\delta[D,P]$ also takes a very simple form, in particular
the quench along the momentum reduces to
$\widetilde{P}^{(U)}_{\mathfrak{p}}=(1+\epsilon)^{-1}
{P}_{\mathfrak{p}}^{(U)}$. Again the traceless condition is
automatically satisfied.
\end{itemize}
Finally, we remark that $A_t=0$ remains solution after the quench.

%%%%%%%%%%%%%%%%%%%%%%%%%%%%%%%%%%%%%%%%%%%%%%%%%%%%%%%%%%%%%%%
\subsection{Features of the Quenched Ensemble}
%%%%%%%%%%%%%%%%%%%%%%%%%%%%%%%%%%%%%%%%%%%%%%%%%%%%%%%%%%%%%%

The outcome of the quench protocol is an initial condition for a new
ensemble $\mathcal{E}'$, which depends explicitly on the {\it quench
  parameters} and implicitly on $h$.  This new ensemble will
evolve on a sub-manifold of the phase space which is not that of
$\mathcal{E}_h$.  As we mentioned, the notion of ``distance'' between
ensembles is better quantified by looking at the variation of the
conserved charges.  For example, if we consider an edge-type quench we
should calculate the expectation value of the following quantities,
\bea
&&\Delta\Tr(X_{\mathfrak{p}} P^I)= \epsilon (P^I_{NN}-P^I_{11})\,, \\
&&\Delta \Tr(P_{\mathfrak{p}} X^I) = \begin{array}{l} \sum_{m\le N}
  \mathrm{Re}(\delta[P_{\mathfrak{p}},D]_{1m} X^I_{m1})\ + \end{array}
\rule{0pt}{.5cm}\nn\\ & & \rule{2cm}{0pt} \begin{array}{l} +\sum_{m<N}
  \mathrm{Re}(\delta[P_{\mathfrak{p}},D]_{Nm}
  X^I_{mN})~. \end{array}\rule{0pt}{.4cm}
\label{quench_diff}
\eea
Since we are especially interested in generating close-to-equilibrium
initial conditions, we will tune the quench parameters in such a way
that the conserved charges admit a perturbative expansion.  Even in
this regime, it is highly non trivial to compute their expectation
value analytically. Nevertheless, simple arguments based on
dimensional analysis and the scaling law~\eqref{scalingBMN} provide us
with the right parametrization in terms of $\epsilon$ and $h$.
Continuing with edge-type quenches, the variation of the energy
$\Delta \langle E\rangle=9h-\HH(\mathcal{E}')$ is then
\bea\label{dym_anal_E}
\frac{ \Delta \langle E\rangle }{9h}& = &1+ \frac{\epsilon}{h^{1/4}} f
\left(\frac{\mathfrak{m}}{h^{1/4}},N \right) + O(\epsilon^2)\,, \\ f&=&
c_0 + c_1 \frac{ \mathfrak{m} }{h^{1/4}} + c_2 \frac{ \mathfrak{m}^2
}{h^{1/2}} + \ldots\,, \nn
\eea
where the $c_i$ only depend on $N$. This is the most general
expression compatible with both the scalings $h\rightarrow \lambda^4
h$, $\epsilon\rightarrow\lambda\epsilon$, and the limit
$\epsilon\rightarrow 0$.  For the BFSS matrix model, the conserved
quantities will only depend on $\epsilon/h^{1/4}$, but in the generic
case there are other contributions in $h$ as long as
$h/\mathfrak{m}\sim 1$.
Arguments based on dimensional analysis are also valid for
inflation-type quenches, the only difference to bear in mind is that
$\overline{\epsilon}$ now plays the role of $\epsilon/h^{1/4}$.  Some
formulae simplify. For example, the order $\overline{\epsilon}$ in the
$\langle E\rangle$ is
\bea
\frac{ \Delta \langle E\rangle }{9h}& = & 1
+\overline{\epsilon}\ \langle \Tr( G_{\mathfrak{p}})\rangle +
O\big(\overline{\epsilon}^2\, \big)\,, \nn\\ G_{\mathfrak{p}}
&=&-P_{\mathfrak{p}}^2 - \sum_J [X_{\mathfrak{p}},X_J]^2 + m^2
X_{\mathfrak{p}}^2\,,
\eea
where $m$ depends on whether $\mathfrak{p}\in \{a,i\}$, according
to~\eqref{dyna_sys_2} or~\eqref{dyna_sys_3}.  $G_{\mathfrak{p}}$
represents the difference between kinetic and potential energy of the
configurations $(P_\mathfrak{p}, X_{\mathfrak{p}})$.  As expected from
considerations about the equipartition theorem, we checked that
$G_{\mathfrak{p}}$ vanishes.  Therefore, $\Delta \langle E\rangle$ is
at least quadratic in $\overline{\epsilon}$ for inflation-type of
quenches.

A more concrete way of understanding how the quench acts on the
ensemble is to consider its consequences on the dynamics of the
observables we are interested in.  According to our previous
discussion, from the knowledge of the distribution of eigenvalues
$\rho^{eig}_I$ of each matrix $X_I$, we completely determine the
expectation value of observables of the type $\OO=\Tr(\MM)$, where
\beq\label{ex_herm_obs_2}
\MM[\{X_I\}]= \sum_{I} c_I f_I(X_I)\,,\qquad c_I\in\mathbb{R}~.
\eeq
Deforming one of these distributions is precisely the task of the
quench protocol.  In fact, by rotating $X_{\mathfrak{p}}$, we obtain
$N$ eigenvalues extracted from $\rho^{eig}_{\mathfrak{p}}$ at the
equilibrium, and by shifting $D$ to $\widetilde{D}$, in practice, we
deform $\rho^{eig}_{\mathfrak{p}}$ by changing its support.
Therefore, a mismatch of the same order of magnitude of the quench
parameters exists between the the shape of the distribution at
$t^+_{quench}$ and that of the equilibrium distribution at
$t^-_{quench}$.  This picture is particularly helpful if we want to
visualize in which way $\mathcal{E}'$ represents a
close-to-equilibrium initial condition from the point of view of the
observable.

\begin{figure}[t]
\includegraphics[scale=0.38]{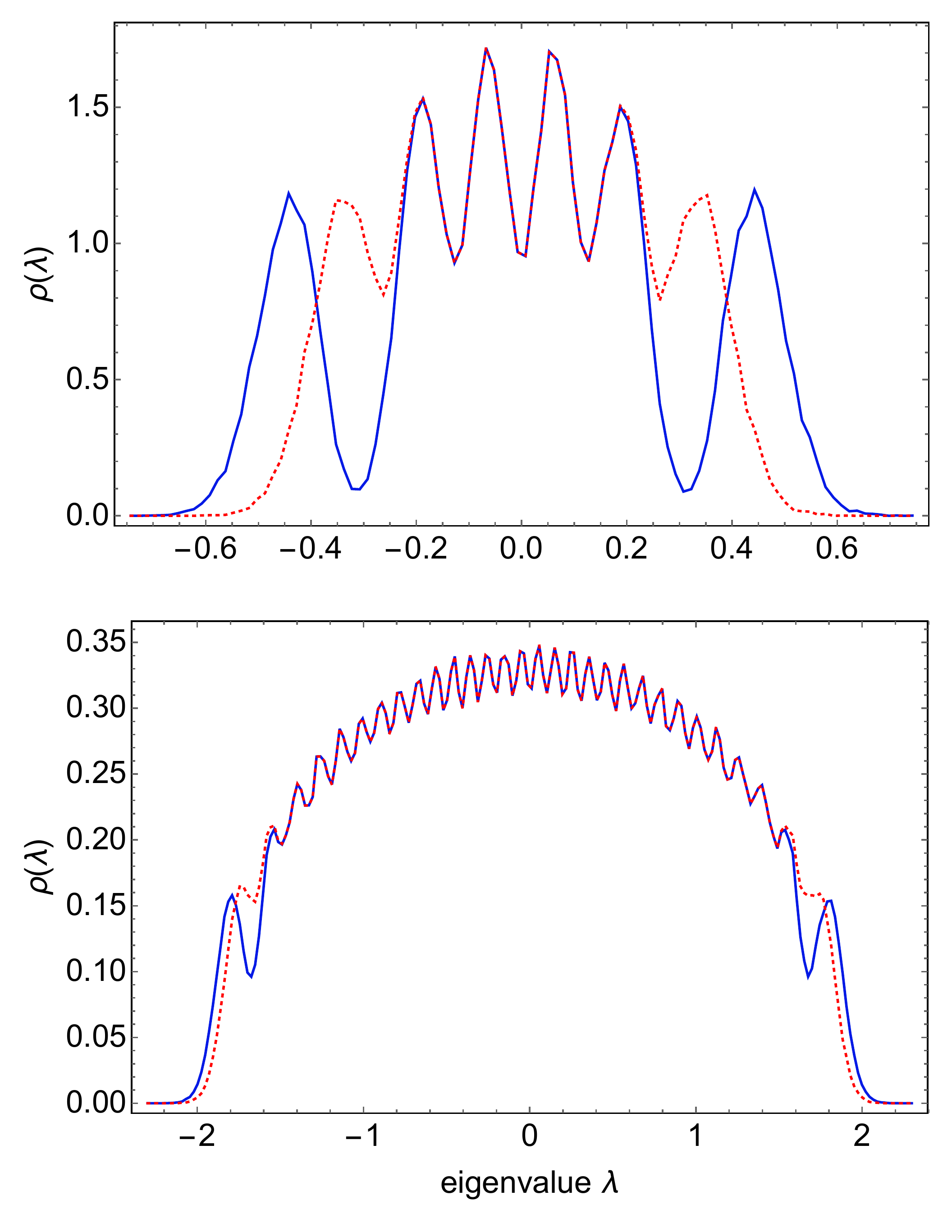}
\vskip -0.2cm
\caption{Top panel: The distribution $\rho^{eig}_{\mathfrak{p}=4}$ at
  $t^{-}_{quench}$ (dotted red curve) and at $t^{+}_{quench}$ (solid
  blue curve), for $N=6$, $h=0.3$ and quench parameter $\epsilon=0.1$.
  Bottom panel: Same as above with $N=30$, $\sigma=1/8$ and
  $\epsilon=0.05$. }
\label{Fig2}
\end{figure}

In Figure~\ref{Fig2} we compare the distribution just before and after
the quench time for edge-type quenches. For any value of $N$, the
equilibrium distribution in $\mathcal{E}$ is characterized by $N$
distinct peaks. For small $N$ the picture is simpler. At
$t^+_{quench}$, we find that the position of the peaks at the
edge of the distribution has moved from the inside out of order
$\epsilon$, whereas the bulk of the distribution has remained almost
unchanged. Since the number of peaks increases with $N$, the same
logic goes through.  We may get a feeling about the large-N result by
repeating the quench protocol in the case of a TGU matrix.  As
Figure~\ref{Fig2} shows, the outcome of the quench produces a ripple
in the equilibrium distribution.

For each event in $\mathcal{E}'$ the $SO(3)\times SO(6)$ symmetry is
broken to the subgroup of rotations that leave $X_{\mathfrak{p}}$
fixed. The breaking is explicit, and the full symmetry is not
restored on the ensemble since the direction of the broken charges is
the same for all configurations. Thus, if $\mathfrak{p}\in\{1,2,3\}$
(or $\mathfrak{p}\in\{4,\ldots,9\}$) $\mathcal{E}'$ preserves
$SO(2)\times SO(6)$ (or $SO(3)\times SO(5)$). After reaching the new
equilibrium in $\mathcal{E}'$, we expect the set of $\rho^{eig}_{I}$
with $I\neq\mathfrak{p}$ to be equal, according to the preserved
symmetries, whereas the distribution $\rho^{eig}_{\mathfrak{p}}$ to be
different in a way which is determined by the strength of the quench
parameter. Given the asymmetry in the eigenvalues
distributions, the charged operators $\NN_{i=1,2}$ will acquire
a non trivial expectation value. 
In the next section we will study the time evolution of these operators 
for close-to-equilibrium quenches. 
Some examples of edge-type quenches in which $\epsilon\gg 1$ 
are illustrated in Appendix {\bf B}.

%==================================================================================
\subsection{Quasi-Normal Modes}
%==================================================================================

\begin{figure}[t]
  \includegraphics[scale=0.32]{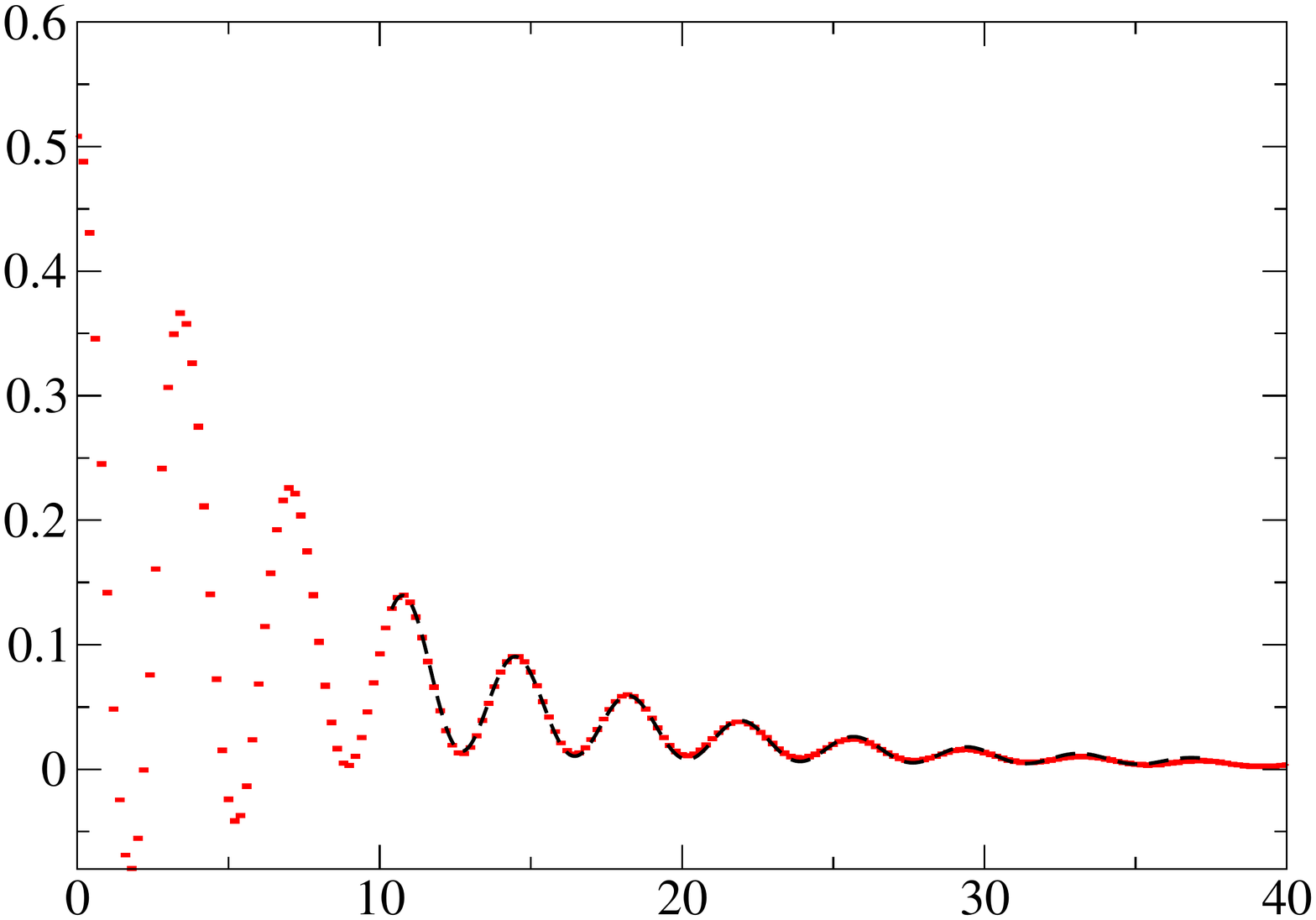}
  \vskip -0.8cm
  \includegraphics[scale=0.32]{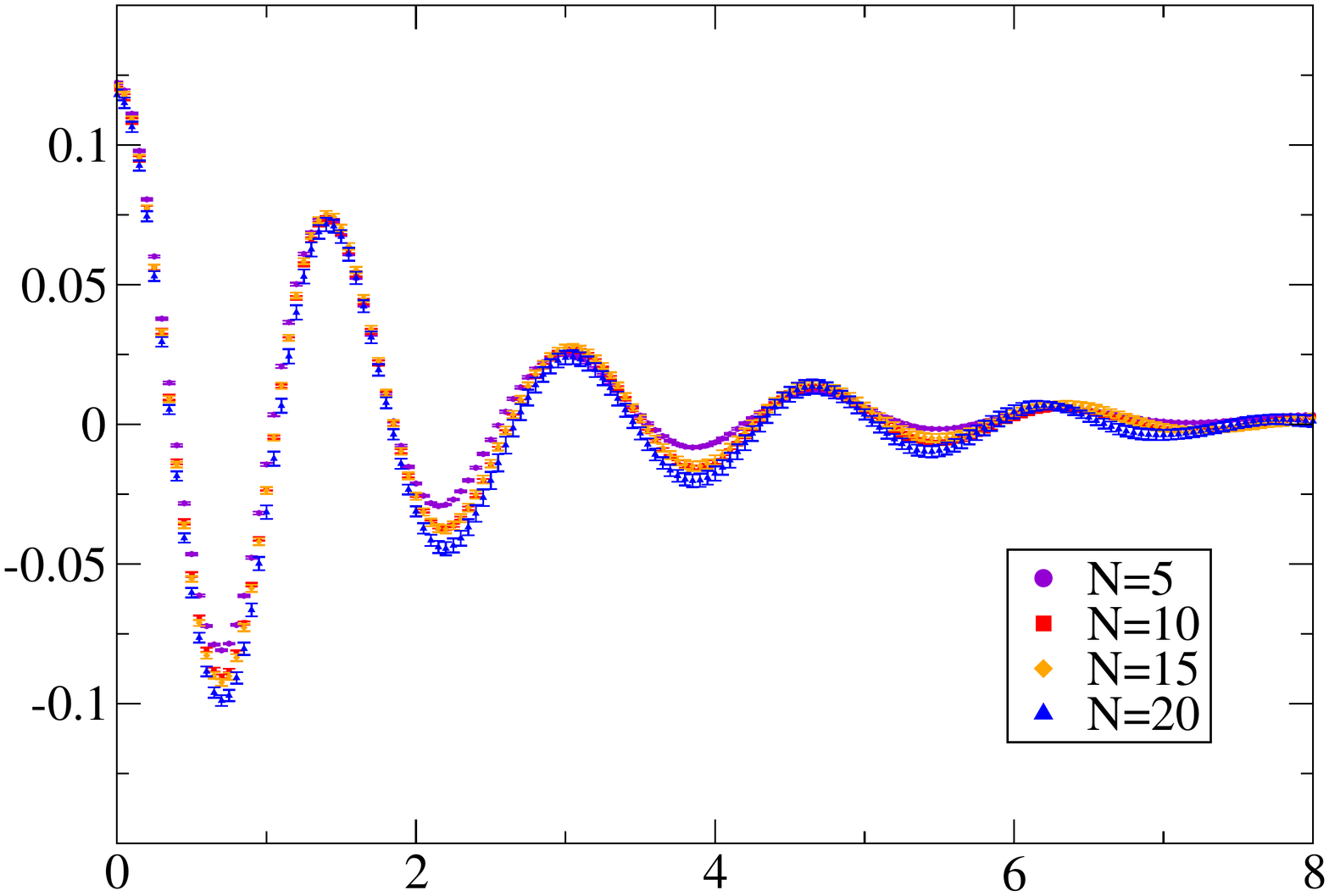}
  \vskip -0.4cm
  \caption{Top panel: The behavior of
    $\langle\NN_2\rangle(t-t_{quench})$ for an edge-type quench, with $h=0.6$, $\epsilon=0.25$, and
    $N=15$.  The solid curve is the lowest quasinormal mode
    obtained from our fit procedure.  Bottom panel: The behavior of
    $\langle\NN_2\rangle(t-t_{quench})$ for an inflation-type quench, various values of $N$,
    $\mathfrak{h}=0.1$, and $\overline{\epsilon} N=0.06$.}
\label{Fig5}
\end{figure}

The time evolution of the observables after the quench has some notable features.
In the top panel of Figure~\ref{Fig5} we have plotted the expectation value of $\NN_2$, 
from the quench to the new equilibration\footnote{We have used a volume of $\mathcal{E}$ of order $10^3$, 
and from each of these configurations at the equilibrium we have evolved approximately $10^3$ quenches.}.
Two facts show up clearly: 
\begin{enumerate}[wide, labelwidth=!, labelindent=0pt]
  \item $\langle \NN_2\rangle$ relaxes fast, without experiencing any non linear transition typical of  the big-bang. 
  \item The process of relaxation is driven by quasi-normal
    oscillations. These modes are collective excitations of the
    non-commutative dynamics.
\end{enumerate}

Before discussing which ansatz describes the quasi-normal ringing, it
is interesting to ask how we should think about these oscillations in
phase space.  Let us consider that at the equilibrium $\mathcal{E}(t)$
is described by a distribution $\rho_t$ which does not change in time,
so intuitively, any realization of $\mathcal{E}(t)$ covers the support
of this distribution with the correct weight.  As we pointed out, the
equilibrium distribution after the quench has moved to a different
support, and evidently $\mathcal{E}'(t^+_{quench})$ does not cover
enough of it.
Under the assumption that $\mathcal{E}'(t^+_{quench})$ is a
close-to-equilibrium initial conditions, we expect the holes between
$\mathcal{E}'(t^+_{quench})$ and the support of the new equilibrium
distribution, to populate quickly.  During this process the dynamics
of the observables is driven by quasi-normal oscillations.  The
ring-down of the observables is in one-to-one correspondence with the
ring-down of the distribution of eigenvalues, since both are functions
of the microscopic degrees of freedom.

The behavior of the $\langle\OO\rangle(t)$ after the quench is a
superposition of quasi-normal modes.  Close to the new equilibrium,
the lowest quasi-normal mode is the dominant one, and the behavior of
the observable is fitted into the following ansatz,
\beq\label{ansatzQNM}
\langle\OO\rangle(t) = \OO_{\infty} + e^{-t/\tau } \Big( d + u
\cos(\Omega t + \phi)\Big)\, ,
\eeq
where $\OO$ stands for any of the observables we are considering. The
details of the fitting procedure are reported in the Appendix {\bf A}.
Here, $\tau^{-1}$ and $\Omega$ are the damping and the ringing
frequencies, respectively.  In complex notation, it is convenient to
define $\omega\equiv\Omega - i \tau^{-1}$.  Tuning the values of $d$ and $u$, the
ansatz accommodates both an over- and an under-damping behavior.  It
is important to emphasize that the quenched ensemble,
by construction, has different properties compared to $\mathcal{E}_h$.
In particular, we expect $d$, $u$ and $\OO_{\infty}$ to be directly
proportionals to the quench parameters because of the explicit symmetry breaking.
Since we are interested in studying the fluctuations of the
equilibrium distribution in $\mathcal{E}_h$, we will consider the limit
$\epsilon\rightarrow 0$, for edge-type quenches, and
$\overline{\epsilon}\rightarrow 0$ for inflation-type quenches.
The only quantities whose limiting values will be non trivial are the
frequencies $\tau^{-1}$ and $\Omega$.

Once again, dimensional analysis provides us with the basic tools to
understand the data.  The complex frequency can be parametrized
as,
\bea
\omega&=& h^{1/4}\Big[ c_0(N) + f\left(
  \frac{\mathfrak{m}}{h^{\,1/4}} ,N\right) \Big]\,, \label{param_omega}
\eea
where $c_0$ is coefficients, and $f$ is a function which vanish in
the asymptotic regime.  The limit $h^{1/4}\gg \mathfrak{m}$ is very
useful here, because as long as the mass parameter is irrelevant the
BFSS and the BMN model have a similar dynamics.  On one hand, we are
free to study the BFSS model independently, setting $\mathfrak{m}=0$,
with the advantage that the model has a simpler dynamics.  On the
other hand, we can check that the values of $\omega$, taken
from the BMN model, asymptote those of BFSS model.
Comparing the results of BFSS simulations at $N=10,15,20$, within error we find the relation 
\beq \label{scaling_reg_BMN}
\omega \propto \ \mathfrak{h}^{\frac{1}{4}}\,,\quad \mathfrak{h}\equiv\frac{h}{N}~.
\eeq
We conclude that in the asymptotic regime, the natural variable upon
which $\omega$ and $\tau$ depend is $h/N$.  We should mention that 
at smaller values of $N$, for example $N=5,6$, we have measured 
small deviations from the scaling regime \eqref{scaling_reg_BMN}. 
On the other hand, the dependence on $\mathfrak{h}$ can be understood as
follows: In the large-N limit we would find the relation
$h/N=\int\rho(\lambda_{\mathcal{H}}) \lambda_{\mathcal{H}}$, where
$\rho$ is the distribution of eigenvalues of the BMN Hamiltonian
$\mathcal{H}$.  Therefore, in order to have a well defined energy
distribution we should keep $h/N$ fixed.

The scaling with $N$ actually holds for the entire profile of the
quasi-normal oscillation.  This is shown in the bottom panel of
Figure~\ref{Fig5}, where we have plotted the behavior of
$\langle\mathcal{N}_2\rangle(t)$ for inflation-type quenches by keeping $\mathfrak{h}$ and $N
\overline{\epsilon}$ fixed.  The latter condition follows from the observation
that
\beq
\langle \Tr\, X_I^2\rangle \propto \sqrt{N h}~.
\eeq
Within error we cannot appreciate any difference on $\langle\mathcal{N}_2\rangle(t)$ among $N=10,15,20$, 
and the deviations visible at $N=5$ can be addressed to $1/N$ corrections.

\begin{figure}[t]
\includegraphics[scale=0.34]{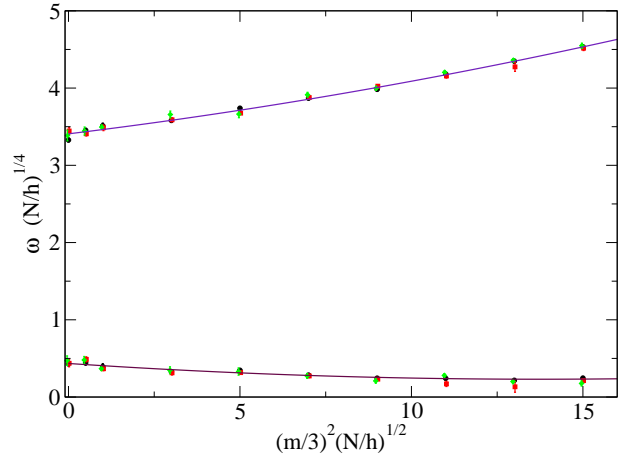}
\vskip -0.3cm
\caption{The curves of $\omega$ for the lowest
  quasinormal mode of $\mathcal{E}_h$, obtained from inflation-type quenches.  
  The parametrization of the
  axis is explained in the main text. The color code is black $N=10$,
  red $N=15$, green $N=20$. (Data have been slightly shifted
  horizontally to improve the readability).}
\label{FigFreqMM}
\end{figure}

Outside the asymptotic regime, corrections due to the mass
$\mathfrak{m}$ cannot be neglected, and we expect them to organize in
a series expansion of the form
\beq
f = c_{1}(N) \left[\, \frac{\mathfrak{m}^4}{ \mathfrak{h}\,
  }\right]^p + {\rm higher\ order\ terms}\,,
\eeq
where $p$ and $c_{1}$ need to be determined. 
We have found that $p=1/2$ and
$c_1$ is, within error, an $N$ independent constant.  In
Figure~\ref{FigFreqMM} we have plotted $\omega \mathfrak{h}^{-1/4}$
as function of
$\mathfrak{m}^2/\mathfrak{h}^{1/2}$.
We can fit these two curves with the ansatz, 
\bea
\omega\, \mathfrak{h}^{-1/4}&=&c_0+ c_1 \left[\, \frac{\mathfrak{m}^4}{ \mathfrak{h}\, }\right]^{{1}/{2}} + c_2 \left[\, \frac{\mathfrak{m}^4}{ \mathfrak{h}\, }\right]\,, \label{fit_num1}
\eea
\rule{0pt}{.3cm}with coefficients: 
\begin{align}
c_{0} &=3.407(10)-i\, 0.435(12)\,,\nn\\
c_{1} & =0.0543(30)+i\,0.0291(32)\,,\nn\\
c_{2}&=0.00138(20)-i\,0.00097(20)~. \label{fitte_num1}
\end{align}
It is worth mentioning that the numerical fit of~\eqref{fit_num1}
do not depend on the details of the quench protocol, in particular we
do not find differences between edge- and inflation-type quenches
when we look at the curves of $\omega\, \mathfrak{h}^{-1/4}$. 
In the linear response regime, this
is a consequence of the fact that the location of the poles of the
Green functions are independent of the strength of the perturbation.

%%%%%%%%%%%%%%%%%%%%%%%%%%%%%%%%%%%%%%%%%%%%%%%%%%%%%%%
\section{V. $\NN=4$ SYM on $\mathbb{R}\times \Sp^3$}
%%%%%%%%%%%%%%%%%%%%%%%%%%%%%%%%%%%%%%%%%%%%%%%%%%%%%%%

The BMN matrix model can be understood as a classical (supersymmetric)
consistent truncation of $\mathcal{N}=4$ SYM on
$\mathbb{R}\times\Sp^3$.  This connection was established
in~\cite{Kim:2003rza}, where the authors showed that the equations of
motions of the $\mathcal{N}=4$ gauge multiplet
$(A_\mu,\chi^{A=1,\ldots,4}_\alpha, \phi^{i=4,\ldots,9})$, truncated
to the lowest harmonics of $\Sp^3$, reduce to those of the BMN matrix
model. The details of the truncation are as follows,
\bea
\phi^i&=&\overline{X}^i(t)\,,\quad A_t=A_t(t)\,,\label{FTscalar}\\ 
A_{\mu=\{\theta,\varphi,\psi\}}&=&
\sum_{c=1}^3 \overline{X}_c(t) {\bf V}^{c+}_\mu({\bf x})\,,\\ 
\chi_\alpha^{A}&=& \sum_{\alpha=1}^2 \theta^A_\beta(t)
{\bf S}^{\beta+}_\alpha({\bf x})\,,
\eea
where $\overline{X}^i \oplus {\bf V}^{c+}\oplus {\bf S}^{\beta+}$ span
the irrep $(1,1,6)\oplus(3,1,1)\oplus(2,1,4)$ of the $SU(2)_+\otimes
SU(2)_-\otimes SU(4)_R$ symmetry group.  In~\eqref{dyna_sys} we
considered only the scalar sector of this truncation, setting the
fermions $\theta^A_\beta$ to zero, and redefining the radius
${\mathsf{R}}$ of the $\Sp^3$ as $\m=6/{\mathsf{R}}$.
Recalling that the covariant derivatives acting on the gauge multiplet
are $D_\mu=\nabla_\mu- i g_{YM} A\wedge$ and $F=dA-i g_{YM} A\wedge
A$, and the quartic coupling is $g^2_{YM}
[\overline{X}_I,\overline{X}_J]^2$, we obtain the BMN action by
considering the field redefinition $\overline{X}= X/g_{YM}$.  The
final result is
\beq
\SS_{\mathcal{N}=4}^{truncated}= \frac{\, vol(\mathbb{S}^3)}{g_{YM}^2}
\SS_{BMN}[\{X_I\},A_t]~.
\eeq
In the regime where $g^2_{YM}\ll 1$ and $\overline{X}
g_{YM}$ is kept fixed, the classical saddle point approximation is justified, and 
the study of the BMN matrix model carried out throughout sections~{\bf
  I}-{\bf IV} can be reinterpreted as the study of $\mathcal{N}=4$ SYM
in the BMN truncation in the classical limit.
The only change we need to implement is to redefine the energy as:
\beq\label{EnergyBMN}
\mathsf{E} = \frac{ vol(\mathbb{S}^3) }{g^2_{YM} }\, \HH\, = \frac{
  9\, vol(\mathbb{S}^3) }{g^2_{YM} }\, h\,,
\eeq
where $\HH$ is the BMN Hamiltonian~\eqref{Hamiltonian}.

In the gauge theory, the Dyson fluid represents a finite energy
ensemble of classical D3-branes in which the off-diagonal modes are
fluctuating. Microscopically, the system is non-commutative and
non-perturbative with respect to the dynamics of the diagonal degrees
of freedom. Therefore, the weak-coupling ``geometric'' interpretation of the
D3-branes has to be rediscovered through the looking glass of the
gauge invariant operators of $\mathcal{N}=4$ SYM. In particular, a
``geometric'' picture should emerge from the description of the 
system offered by the eigenvalue distributions
associated to each single trace operators.
This picture is inspired by the AdS/CFT correspondence which, 
in the regime of strong 't Hooft coupling, relates the dynamics of 
field configurations to those of a gravitational problem in the Anti-de-Sitter space. 
Being aware that a direct comparison between AdS physics and our Dyson fluid would not be possible, 
we find important to study the two dynamics in a non trivial case, precisely with the aim of highlighting the major differences.

In the following we review basic aspects of the AdS/CFT correspondence, 
and we will compute quasi-normal modes for the dual operators 
that we studied in the BMN truncation. Since at this point we are only interested 
in qualitative features, we will carry out our toy model computation 
in a simple black-hole geometry, and we will not go beyond the AdS$_5$-Schwarzschild black hole.

%%%%%%%%%%%%%%%%%%%%%%%%%%%%%%%%%%%%%%%%%%%%%%%%%%%%%%%%
\subsection{AdS/CFT Correspondence and Black Holes}
%%%%%%%%%%%%%%%%%%%%%%%%%%%%%%%%%%%%%%%%%%%%%%%%%%%%%%%%

One of the greatest achievements of string theory has been the
discovery of the AdS/CFT correspondence, or more generically, of the
gauge/gravity dualities.  For $\mathcal{N}=4$ SYM the dual space-time is an
asymptotically AdS$_5\times \Sp^5$ background.  The two sides of the
duality are related by the the D-brane construction of the field
theory~\cite{Maldacena:1997re} which sets,
\beq
L_{\Sp^5}^4=L_{AdS_5}^4 = 4\pi\alpha^2 (g_s N)\,,\qquad g_s=g^2_{YM}\,,
\eeq
where $L_{AdS_5}$ and $L_{S^5}$ are the radii of AdS$_5\times \Sp^5$,
$g_s$ is the dimensionless string coupling, and $\alpha$ is the string
tension. The radii of AdS$_5$ and $S^5$ are equal and we shall refer
to them simply as $L$. The reader unfamiliar with the AdS space might
find useful to think about it as a ``box'' of constant negative
curvature proportional to $1/L^2$.

The quantity $L^2/\alpha$ depends only on the 't Hooft coupling
$\lambda_t\equiv g_{YM}^2 N$.  Keeping $\lambda_t$ fixed, and taking
the large-N limit, the string theory becomes perturbative,
i.e. $g_s\ll 1$. Then, two cases are well under control:
\begin{enumerate}[wide, labelwidth=!, labelindent=0pt]
\item
  When $\lambda_t \gg 1$, the $AdS_5\times \Sp^5$ background is
  classical and the string theory can be truncated to $10d$ classical
  gravity with Newton constant $G_{10}= 8\pi^6
  g_s^2\alpha^4$~\cite{D'Hoker:2002aw}. In this regime, the duality
  is very powerful and predicts that $\mathcal{N}=4$ SYM at strong
  coupling is dual to a gravitational theory in $AdS_5\times \Sp^5$.
  The holographic dictionary provides the concrete link between the
  two theories~\cite{Witten:1998qj}. In particular, the $4d$
  space-time, where the field theory lives, is identified with the
  boundary of AdS$_5$, and {\it chiral single-trace} operators in the
  field theory are mapped to bulk fields of the $10d$ geometry.
  Intuitively, any field configuration at the boundary now acquires a
  bulk profile along the the fifth (extra) {\it radial} coordinate of
  AdS$_5$. This bulk profile is obtained by solving classical
  gravitational equations of motion. The near-boundary behavior of
  bulk fields plays an important role in the holographic dictionary,
  and a more precise statement about it will be made in the next
  section.
\item When $\lambda_t\ll 1$ the field theory is perturbative.  In this
  regime the curvature of the gravitational background is large, and
  the classical spacetime structure has to be modified by quantum gravity corrections.
\end{enumerate}

It is important to realize that according to the basic principles of
holography, asymptotically AdS solutions are specified uniquely by the
set of conserved charges of the boundary field theory, together with
the expectation values of charged operators.  These are the same
charges we used to initialize and characterize $\mathcal{E}$, as well
as more complicated ensembles like $\mathcal{E}'$. Therefore,
in order to set up a comparison with $\mathcal{E}$, we should at least look for
a gravity solution with finite energy, 
zero momentum\footnote{ It is simple to show that, upon using the gauge constraint, 
the $T_{t\mu=\theta,\varphi,\psi}$ components of the stress energy tensor of 
$\mathcal{N}=4$ SYM in the BMN truncation, 
are proportional to ${\bf V}^{a+}_\mu L_c$, where $L^c$ are the $SO(3)$ charges \eqref{Noether_Charges}.
Therefore $T_{t\mu=\theta,\varphi,\psi}$ vanish on $\mathcal{E}_h$.}, 
and zero $SO(6)$ charges. 
Moreover, we should also look for a solution which
supports quasinormal oscillations. Together, these two observations
point to the simplest and most primitive candidate: the
AdS$_5$-Schwarzschild black hole.

Black holes have special properties: Since the seminal works
of Bekenstein, Hawking and Gibbons, it has been recognized that
thermodynamic variables can be assigned to stationary black holes~\cite{Bekenstein:1973ur,Gibbons:1976ue}, and via the
AdS/CFT duality it has been understood that an asymptotically AdS
black hole corresponds to a field configuration
which is approximately thermal.  Black holes do not only appear as
static objects, but they are also characterized by important dynamical
features, in particular by the spectrum of their quasinormal
oscillations.  In this respect, let us notice that AdS$_5$ would not have worked 
for our comparison, simply because at the linearized level it only allows for normal
oscillations.  By the AdS/CFT duality, the ringing frequency and the
decay rate of the lowest quasinormal modes of the black hole control
the scales of the process of ``thermalization'' in the field
theory~\cite{Berti:2009kk}.

%%%%%%%%%%%%%%%%%%%%%%%%%%%%%%%%%%%%%%%%%%%%%%%%%%%%%%%%
\subsection{The AdS$_5$-Schwarzschild black hole} 
%%%%%%%%%%%%%%%%%%%%%%%%%%%%%%%%%%%%%%%%%%%%%%%%%%%%%%%%

In our notation the AdS$_5$-Schwarzschild black hole is described by
the metric
\bea\label{coordScW}
ds^2_{BH} &=& - f(r) dt^2 + {r^2}\, R^2d\Omega_3^2 +
\frac{dr^2}{f(r)}\,,\\ \rule{0pt}{.5cm} f(r)&=& \frac{r^2}{L^2} +
\frac{1}{R^2} - \frac{M L^2 }{r^2}~.
\eea
This background is a solution of the $5d$ Einstein action,
\bea
I=\frac{1}{16\pi G_5 } \int\sqrt{g}\ (\RR - \Lambda) + I_{GH}\,,
\eea
where $I_{GH}$ is the standard Gibbons-Hawking surface term, and
\bea\label{expr_G5}
\Lambda= -\frac{12}{L^2}\,,\qquad \frac{1}{G_5}=\frac{ 2 N^2}{\pi
  L^3}~.
\eea
The $5d$ Newton constant $G_5$ is obtained as the ratio of $G_{10}$
over the volume of the ${\Sp^5}$, which is trivially fibered in the
full 10$d$ geometry of string theory.  $G_{10}$ has a stringy
expression in terms of $\alpha$~\cite{D'Hoker:2002aw}, thus upon
substituting $L$ for $\alpha$, we obtain $G_5$ as written
in~\eqref{expr_G5}.
 
The radial coordinate $r$ extends from the boundary at infinity,
corresponding to $\mathbb{R}\times \Sp^3$, to the radius of the
horizon, hereafter $r_h$, which is defined as the greatest root of the
equation
\beq\label{horizon}
f(r_h)=\frac{r_h^2}{L^2} + \frac{1}{R^2} - \frac{M L^2 }{r_h^2}=0~.
\eeq
The parameter $M$ is called the {\it non-extremality parameter}. When
$M=0$ the geometry is that of pure $AdS_5$ in global coordinates.  The
radius of the boundary $\Sp^3$ is measured in units of $L$,
i.e. $R=\mathsf{R}/L$, where $\mathsf{R}$ is the field theory
quantity.

%=============================================================================
\subsection{Boundary Energy and Thermodynamics} 
%=============================================================================

As we mentioned earlier, the mapping between gravitational solutions
and field theory configurations goes through the correct
identification of the conserved charges on both sides. In our case,
the energy.  From the bulk perspective, the quantity we need to
evaluate is the time component of the stress-energy tensor integrated
over the three-sphere at the boundary. This is properly defined as
\beq\label{T_tt_bulk}
\mathsf{E}=\int d\Omega_3\, \sqrt{g}\Big|_{\mathbb{S}^3} (u^\mu\,
T^{reg}_{\mu\nu}\, \xi^\nu)\ \Big|_{r=\infty}\,,
\eeq
where $u^\mu=(u^t,0,0,0,0) $ is the time-like unit vector orthogonal
to $\mathbb{S}^3$, and $\xi^\nu$ is the Killing vector
$\sqrt{g_{tt}}\, u^\mu$.  The stress tensor is
\beq
T^{ab}= \frac{2}{\sqrt{-h}}\frac{\delta I}{\delta h_{ab}}\,,
\eeq
where $h_{ab}$ is defined from writing the metric as $ds^2_5=
\mathscr{N}^2 dr^2 + h_{ab} (d\sigma^a+ \mathscr{V}^a dr)(d\sigma^b+
\mathscr{V}^b dr)$ (see~\cite{Liu:2003px} for our conventions).  The
expression of $T^{reg}_{\mu\nu}$ includes counter-terms which regulate
well understood divergences in AdS. The result for $T^{reg}_{\mu\nu}$
is
\bea
T_{tt}^{reg}&=&\frac{1}{8\pi G_5} \frac{3L}{2r^2} \left( M +
\frac{1}{4 R^4} \right)\,,
\eea
Plugging this expression in \eqref{T_tt_bulk} and using the relation
between $G_5$ and $L$, we finally obtain
\bea
{ \mathsf{E} } &=& \frac{3\, vol(\mathbb{S}^3)}{8 \pi^2} {} \left(
\frac{N^2 M}{L^4}+ \frac{N^2}{4 { \mathsf{R}} ^4} \right)\,,
\eea
where $vol(\mathbb{S}^3)=2\pi^2 { \mathsf{R}} ^3$. The answer for
$\mathsf{E}$ is the sum of a zero point (Casimir) energy proportional
to $1/{ \mathsf{R}}^4$ and a physical $M$-dependent energy which can then be
rewritten in the form,
\bea\label{EnergyBH}
{ \mathsf{E}_{phys.} } &\equiv& \frac{3\, vol(\mathbb{S}^3)}{8 \pi^2}
{} \frac{N M}{4\pi g^2_{YM} }~.
\eea
The factors of $g_{YM}^2$ and ${vol(\mathbb{S}^3)}$ in this formula
are the same as in the field theory definition of the energy, and the
actual value of the energy is proportional to $N$ times $M$.  In the
field theory ${ \mathsf{E} }$ is computed thought the path integral
with the insertion of the Hamiltonian, which is a single trace
operator.  Therefore, ${ \mathsf{E} }$ will be proportional to $N$
times the integrated distribution of the eigenvalues of the effective
Hamiltonian at strong 't Hooft coupling, i.e. $M$.  We recover in this
way, the original meaning of the black hole parameter $M$, which was
first derived in the seminal paper~\cite{Malda}.

The energy $\mathsf{E}$ can also be interpreted as {\it thermal} by
introducing the Hawking temperature,
\bea
T_H &=& \frac{1}{4\pi} f'(r_h)= \frac{1}{\pi L} \left( \frac{r_h}{L} +
\frac{ L/R^2 }{2 {r}_h} \right)~.
\eea
In this formulation, the partition function is given by the the
on-shell value of the (regularized) gravitational action,
\beq\nn
I_{reg} =\beta_H \frac{\, vol(\mathbb{S}^3) }{16\pi\, G_5L} \left(
\frac{ {r}_h^{\,2} /R^2}{L^2} - \frac{ {r}_h^{\,4}}{L^4} +\frac{3}{4
  R^4} \right)\,,
\eeq
the thermal energy $E_{th}$, and the Bekenstein-Hawking entropy
$S_{BH}$ can be obtained as follows,
\bea
E_{th}&=&\frac{\partial I_{reg}} {\partial \beta_H } =\frac{3\,
  vol(\mathbb{S}^3) }{16 \pi\, G_5 L} \left( \frac{{r}_h^{\,4} }{L^4}
+ \frac{ {r}_h^{\,2}/R^2 }{L^2} +\frac{1}{4 R^4} \right)\,,
\nn\\ S_{BH}&=& \beta_H E - I = \frac{ vol(\mathbb{S}^3) }{4 G_5 L^3}
\, r_h^3 =\frac{A(Horizon)}{4 G_5}~.
\eea
It is simple to check from \eqref{horizon} that $E_{th}=\mathsf{E}$.

For a given temperature $T$ there exist two solutions of the equation
$T_H(r_h)=T$, and the corresponding black holes are dubbed as
``large'' if $R\, {r}_h/L> 1/\sqrt{2}$, and ``small'' if $R\, {r}_h/L<
1/\sqrt{2}$.  Large black holes have positive specific heat and are
dual to thermal states in the field theory.  Small black holes are
always thermodynamically irrelevant since an Hawking-Page transition
between large black holes and thermal AdS takes place at $R\,
\overline{r}_h =1$.  In the field theory, this Hawking-Page transition
has been interpreted as a second order
transition~\cite{Witten:1998zw}.

%%%%%%%%%%%%%%%%%%%%%%%%%%%%%%%%%%%%%%%%%%%%%%%%%%%%%%%
\subsection{Massive Quasinormal Modes\\ of the AdS$_5$-Schwarzschild black hole}
%%%%%%%%%%%%%%%%%%%%%%%%%%%%%%%%%%%%%%%%%%%%%%%%%%%%%%%

In this section we study bulk quasinormal modes of the
AdS$_5$-Schwarzschild black hole which are dual to the scalar
operators in the representation ${\bf 20}$ of $SO(6)$.  Let us recall
that under $SU(3)\subset SO(6)$ the representation ${\bf 20}$
decomposes into ${\bf 8}\oplus {\bf 6}\oplus {\bf \bar{6}}$ and
contains the operators $\mathcal{N}_{i=1,2}$ and
$\mathcal{C}_{i=4,\ldots 9}$ that were analyzed earlier in the context
of the BMN matrix model.

On the gravity side, the $SO(6)$ R-symmetry is realized geometrically
as isometries of the $\Sp^5$.  Each scalar operator sitting in a
representation of $SO(6)$ is mapped to a specific harmonic deformation
of the $\Sp^5$.  The harmonics of the $\Sp^5$ sitting in the
representation ${\bf 20}$ are charged under the isometries of the
$5$-sphere, and therefore must be coupled to bulk gauge fields.  These
bulk fields sit in the irrep ${\bf 15}$ of $SO(6)$, and their
time-components are in one-to-one correspondence with the boundary
charges $J_{q=1,\ldots 15}$.  It is perhaps useful to sketch how these
fields are realized in concrete as deformations of the $\Sp^5$.
Following the notation of~\cite{Cvetic:2000nc}, the ${\bf 20}$ is
parametrized by the matrix $T_{ij}=T_{ji}$ in $SL(6,\mathbb{R})$, and
the metric of the $\Sp^5$ is written as
\beq
ds^2_{\Sp^5} =\, \delta^{ {-} \frac{1}{2}}\, D\mu^i\ T^{-1}_{ij}\,
D\mu^j\,,\qquad \delta= \mu^i\, T_{ij}\, \mu^j~.
\eeq
The coordinates $\mu^{i=1,\ldots 6}$ are subject to the constraint
$\mu^i\mu^i=1$, and the $1$-forms $D\mu^i$ are the covariant
derivatives $D\mu^i= d\mu^i + A^{ij} \mu^j$, where the matrix of bulk
gauge fields $A^{ij}=-A^{ji}$ represents the ${\bf 15}$.  The
covariant derivative on $T$ is $DT=dT + [A,T]$. 
The round $\Sp^5$ is recovered from the 
trivial configuration $T={\rm diag} (1,1,1,1,1,1)$ with all
gauge fields turned off. Two are the
$5d$ solutions in which $T$ can be taken to be trivial: empty AdS$_5$, and the
AdS$_5$-Schwarzschild black hole.
The authors of~\cite{Cvetic:2000nc} found out the fully non-linear
consistent truncation of type IIB supergravity, which only retains
gravity and the fields in the ${\bf 20} \oplus {\bf 15}$.  Any charged
black hole in this sector, bold or hairy, can in principle be
constructed.  For hairy black holes, the field theory configuration is
further characterized by the expectation value of the corresponding
operators in the ${\bf 20}$. Turning on expectation values for
operators of the type $C_{i_1,\ldots i_n} Tr(X^{i_1}\ldots X^{i_n})$,
with $n>2$ and $C_{i_1,\ldots i_n}$ totally symmetric and traceless,
cannot be done within consistent truncation ans${\rm\ddot{a}}$tze, but
requires non separable $10d$ gravitational backgrounds.  Coulomb
branch solutions are an example,~\cite{Skenderis:2006di}.

In general, quasinormal modes materialize as linearized perturbations
of black hole backgrounds.  For the AdS$_5$-Schwarzschild black hole,
we are interested in perturbing the $T_{ij}$ sector. Setting
$T_{ij}=\delta_{ij} + \epsilon\, \Phi_{ij}$, we truncate the equations
of motion at order $\epsilon$.  The $\Phi_{ij}$ decouple and each
component satisfies the equation,
\beq\label{eqmotio1}
\frac{1}{\sqrt{g}} \partial_\mu ( \sqrt{g} g^{\mu\nu}\partial_\mu \Phi
) = m^2 \Phi\,,
\eeq
where $m^2=-4/L^2$.  According to the holographic dictionary we expect
$\Delta(\Delta-4)=m^2L^2$, where $\Delta$ is the dimension of the dual
operator\footnote{
There cannot be masses such that $\Delta$ is less than the unitarity bound. 
The bound for $m^2L^2$ corresponds to the celebrated Breitenlohner-Freedman bound in AdS.}.
For the case at hand $\Delta=2$

We shall look for solutions of~\eqref{eqmotio1} with the following profile,
\beq
\Phi=e^{-i\omega t }\, \phi(r)\, Y_0^{(0,0)}(\, \Omega_3)\,,\qquad
Y^{(0,0)}_0=1~.
\eeq
Different harmonics of the spatial $\Sp^3$ could also be considered,
and the calculations would go through with minor modifications.  We
have taken the lowest harmonic $Y_0^{(0,0)}$ so to
reproduce~\eqref{FTscalar} at the boundary.
Changing variables to $z=L^2/r^2$ simplifies the equation of motion to
\beq\label{eqbetter}
\phi'' +\left( \frac{b'(z)}{ b(z)} -\frac{1}{z}\right)\phi' -
\frac{1}{4}\frac{m^2L^2}{z^2 b(z)} +\frac{1}{4}\frac{\omega^2L^2}{z\,
  b(z)^2}=0\,,
\eeq
where $b(z)=1 + z/R^2 - M z^2$. In this new coordinate $z$, the
boundary is placed at $z=0$ and the horizon is
\beq
z_h= R^2 \ \frac{ 1+ \sqrt{1+ 4 x} }{2x}\,, \qquad {x}\equiv M R^4~.
\eeq
The near-boundary behavior of $\phi$ is
\beq
\phi(z\rightarrow 0)= A\, z + B\, z \log z\,,
\eeq
with $A$ and $B$ constants. By rescaling $z\rightarrow R^2 \bar{z}$,
it is simple to show that $\mathfrak{w}\equiv\omega R L$, and $x=
MR^4$ are the only parameters entering the equation of $\phi$.  By the
holographic dictionary, $A$ will be interpreted as being proportional
to $\langle \OO\rangle$, whereas the $\log z$ term will be interpreted
as a source in the field theory~\cite{Bianchi:2001kw}.  Here $\OO$ is
any operator in the ${\bf 20}$, since as we mentioned, at the
linearized level the $\Phi_{ij}$ decouple.

We are interested in studying how the perturbation of the black
hole relaxes when no boundary sources are turned on\footnote{
From the point of view of $\mathcal{N}=4$ SYM on $\mathbb{R}\times \Sp^3$, 
the mass terms due to the curvature of the sphere are effectively sources for $\OO_s^{(3)}$ and $\OO_s^{(6)}$. 
However, these two operators are non-chiral  and do not appear in the spectrum of supergravity.}.   
Then, we shall find solutions of~\eqref{eqbetter} such that
$B(\mathfrak{w},x)=0$. In order to properly
define the Cauchy problem for $\Phi$, we also have to specify a
``regularity'' condition at the horizon.  Near the horizon, a generic
solution would behave like
\bea
\Phi(\bar{z}\rightarrow \bar{z}_h)&=& e^{-i\omega t} (
\bar{z}_h-\bar{z} )^{\,\pm\, i\, \mathfrak{w} / \mathfrak{u}
}\,,\\ \mathfrak{u}&=&-b'(\bar{z}_h)\bar{z}_h^{\frac{1}{2} } >0\,,
\eea
where the $+(-)$ sign corresponds to out-going (in-going) modes.
Classically, matter can only fall into the horizon, therefore the only
allowed solution has to be in-going at the horizon.  With this second
boundary condition the Cauchy problem is well defined.  As shown in
the pioneering work~\cite{Horowitz:1999jd}, it is a general fact that
the condition $B(\mathfrak{w},x)=0$, for fixed $x$, admits a discrete
set of complex solutions of the form $\mathfrak{w}=\mathfrak{w}_{\rm
  Im} - i \mathfrak{w}_{\rm Re}$ with $\mathfrak{w}_{\rm Re}>0$.
These are the so called {\it quasi-normal frequencies}, and the lowest
one characterizes how the perturbation equilibrates at late
times.  As it should be, the decay rate $\mathfrak{w}_{\rm Re}$ comes
out positive.

\begin{figure}[t]
\includegraphics[scale=0.51]{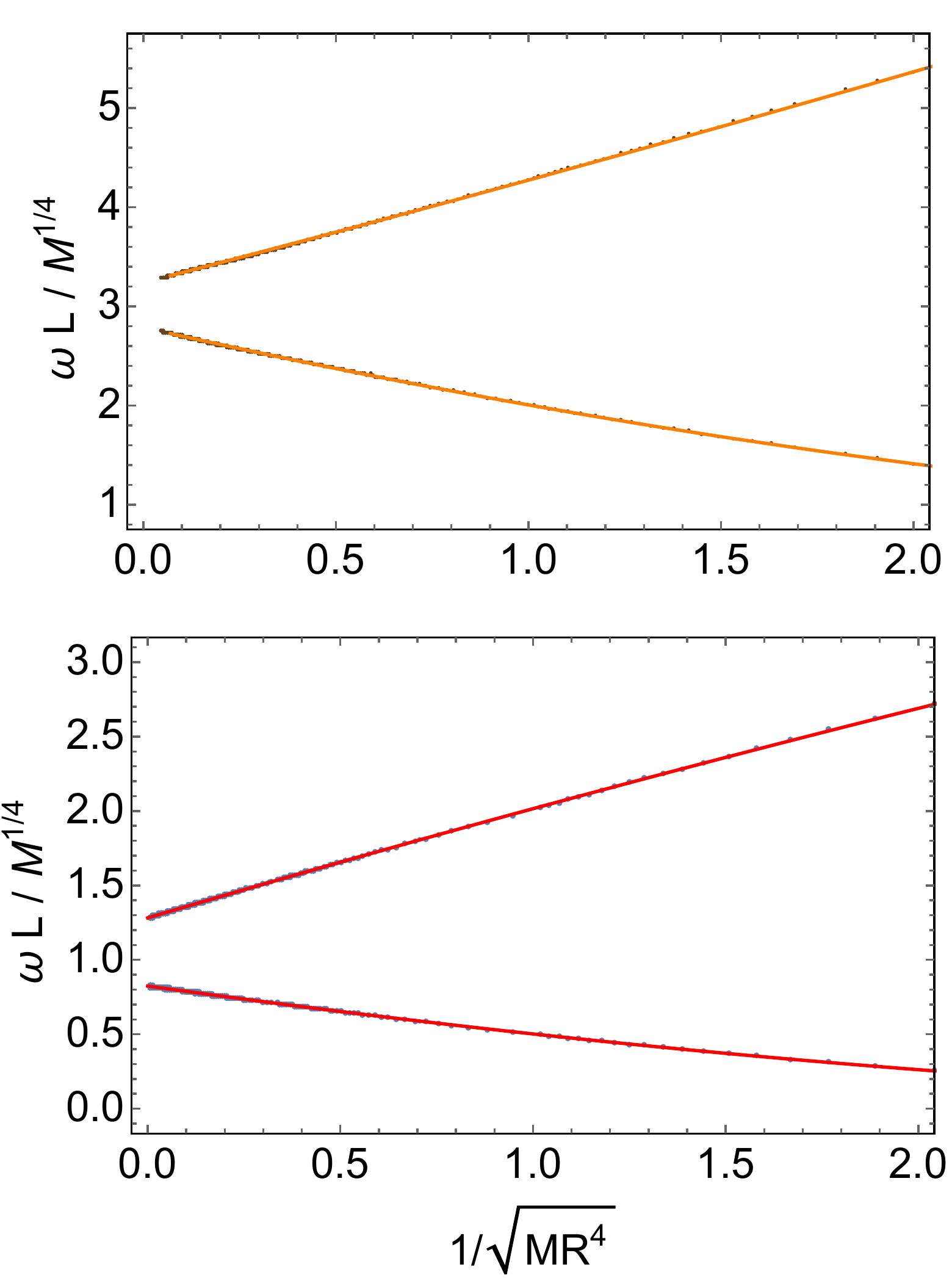}
\vskip -0.1cm
\caption{Lowest (bottom panel) and next-to-lowest (top panel)
  quasinormal frequencies for the scalar perturbation $\phi$. The solid (red and orange) lines are the fit.}
\label{FigQNM_BH}
\end{figure}

We have solved the equation of motion for $\phi$ numerically,
interpolating between a series solution at the boundary and a series
solution at the horizon.  Scanning through the complex frequency plane
$\mathfrak{w}$, we have found the quasinormal frequencies as function
of $x$, i.e we have found the curve
$\mathfrak{w}=\mathfrak{w}(x)$.  It is interesting to recover
from this result the relation between $\omega$ and the scales of the
problem, $M$ and $R$, according to dimensional analysis arguments.  In
fact, it is simple to show that in the limit of large $R$ there is only one possibility, 
$\mathfrak{w}(x)= c_{0} x^{1/4}$, and therefore $\omega L = c_{0}
M^{1/4}$.  The finite $R$ dependence instead is non trivial, 
and follows from the actual numerical solutions. 
The result fit into the ansatz,
\beq\label{freq_BH_fit}
\omega L = M^{1/4} \left( c_{0} + \frac{c_{1}}{x^{1/2}} +
\frac{c_{2}}{x} + \ldots \right)\,,
\eeq
where $\omega L$ is the proper frequency at the boundary.  The
numerical results for the lowest and the next-to-lowest quasi-normal
modes are shown in Figure~\ref{FigQNM_BH}.  
For the lowest quasi-normal mode, the values of the first
two coefficients in the fit ansatz \eqref{freq_BH_fit} are,  
\begin{align}  
c_{0}  &=\ 1.282-i\,0.824\,,\nn\\
c_{1}  &=\ 0.763+i\,0.362\,,\nn\\ 
c_{2} &= -0.029- i\, 0.041\,,\label{fitte_num2}
\end{align}
and for next-to-lowest quasi-normal mode,  
\begin{align}  
c_{0}  &=\ 3.241-i\,2.786\,,\nn\\
c_{1}  &=\ 0.985+i\,0.873\,,\nn\\
c_{2}& =\ 0.045- i 0.093~.
\end{align}

%%%%%%%%%%%%%%%%%%%%%%%%%%%%%%%%%%%%%%%%%%%%%%%%%%%%%%%
\section{VI.\, Discussion}
%%%%%%%%%%%%%%%%%%%%%%%%%%%%%%%%%%%%%%%%%%%%%%%%%%%%%%%

Connecting the dots of what we have done so far, we have two parallel situations in $\mathcal{N}=4$ SYM, a statistical ensemble
on one side, a black hole on the other side, and two dynamical quantities which we can now compare: their lowest quasinormal frequency. 
%
%playing a role
%in the framework of the AdS/CFT correspondence, and both fluctuating
%through quasinormal oscillations. A comparison between them is inevitable.  
%
Let us here emphasize that there is a-priori no
reason why we should expect some kind of resemblance, but according to our
findings, their qualitative behavior is surprisingly similar. 
The parameters, $\mathfrak{h}$ and $\mathfrak{m}$ in the BMN
matrix model, and $M$ and $R$ in the black hole, determine the
frequencies very much in the same way. In particular, it is an
independent outcome of the two calculations that the leading
correction to the flat or zero mass limit, written as an expansion in
powers of $\mathfrak{m}^4/\mathfrak{h}$ and $1/MR^4$, is fixed by the
power $p=1/2$. This result cannot be obtained by dimensional
analysis. %Moreover, the $N$ dependence in $\mathfrak{h}=h/N$ agrees
%with that of the black hole and can be inferred by comparing the
%expressions of the energy in~\eqref{EnergyBMN} and~\eqref{EnergyBH}.
In the absence of a better analytical understanding, we can only appreciate 
the unexpected beauty of the precision fit
in Figure~\ref{FigFreqMM} while considering its striking similarity with that of Figure~\ref{FigQNM_BH}.

%As the Dyson fluid and the black hole belong to different regimes, 
%the correspondence observed is obviously not exact.
%For example, after using the
%information coming from dimensional analysis, undetermined numerical
%coefficients can only be function of the
%coupling, whose value is effectively different in the Dyson fluid
%and in the black hole. As expected the coefficients of the best fit of the ansatz \eqref{fit_num1} 
%in the BMN model, listed in \eqref{fitte_num1}, do not coincide with the ones obtained from the gravitational fit in \eqref{fitte_num2}.
%

%From the gauge theory viewpoint, the black hole is a quantum field
%configuration whose energy has finite expectation value.  
Following the work of~\cite{Berkowitz:2016jlq,Catterall:2014vka}, it would be interesting to
compute semiclassical corrections to our Dyson fluid distribution, 
and see how the quasi-normal frequencies are modified at small but finite 't Hooft coupling.  
%since as the coupling is increased ``virtual''
%configurations will start contributing to the path integral.  
%Then, the
%path integral computation will provide the understanding of both the
%coupling dependence of the quasi-normal frequencies and the mapping of
%the Dyson fluid onto the black hole ensemble. In particular, we might
%be able to observe a similar $L$ dependence, as in~\eqref{freq_BH_fit}, and what
%fraction of the mass of the black hole corresponds to $h/N$ in the
%classical limit.

%%%%%%%%%%%%%%%%%%%%%%%%%%%%%%%%%%%%%%%%%%%%%%%%%%%%%%%
\section{VII.\, Conclusions and Outlook}
%%%%%%%%%%%%%%%%%%%%%%%%%%%%%%%%%%%%%%%%%%%%%%%%%%%%%%%

In this paper we have analyzed Dyson fluid solutions of the BMN matrix model
living on Lagrangian sub-manifolds of the phase space.  
We have characterized the corresponding ensemble by symmetries, 
conserved charges, and
at the equilibrium, by the expectation values of local operators.
We have defined a novel gauge-invariant quench protocol 
which allowed us to deform the equilibrium distribution in a controlled way, and we have
shown that the expectation values of twist two scalar operators 
in the $SO(6)$ sector re-equilibrate via quasi-normal oscillations.
Finally, we have determined the numerical dependence of the complex frequency
$\omega$ of the lowest quasi-normal mode 
as function of the energy $\mathfrak{h}$ 
and the dimensionless parameter $\mathfrak{m}^4/\mathfrak{h}$. 

The interesting features of the Dyson-fluid are 
triggered by the non-commutative nature of the microscopic dynamics.
%Nevertheless, 
The study of eigenvalue distributions of observables
offers an alternative geometric picture of the dynamics.
In particular, 
the complexity of an observable provides the tool 
to detect different type of correlations in the ensemble. 
%
%The observables cannot be identified completely within the matrix model 
%and the correct identification follows instead from the underlying gauge theory, i.e. $\mathcal{N}=4$ SYM on $\mathbb{R}\times \mathbb{S}^3$.
In the second part of the paper we have compared   
the emergent geometric structure of our classical Dyson-fluid
and the holographic geometry of a black hole at strong 't Hooft coupling. 
Our main result has been to point out an unexpected similarity 
between the parametric behavior of the quasi-normal frequencies 
in the Dyson-fluid and in the black hole background.  
%In a future work we will extend this comparison to spin-2 waves  
%by considering on one side the BMN graviton, and on the other side a gravitational wave.  
%

There are similarities between the phase space picture of the quasi-normal
oscillations, and the entropic principle of \cite{Verlinde:2010hp}: 
A rigorous way to understand the process of equilibration would be 
to consider a covering of the allowed phase space and verify that 
the population in each patch does not change in time. 
In this language, we are then led to conjecture the existence of an 
effective actions whose equations of motions determine the equilibrium distributions of the operators. 
As it happens for the case of
the Brownian motion \cite{Blaizot:2009ex}, 
it would be interesting to rewrite the time evolution 
of the distributions in terms of fluid-like equations.

\vspace{0.5cm}

%%%%%%%%%%%%%%%%%%%%%
\begin{acknowledgments}
\paragraph{Acknowledgments.}
We are in debt with: Diego Hofman, for valuable discussions and
important comments on the draft at different stages of this work,
Dario Villamaina, for collaboration at an early stage, and Vasilis
Niarchos for important feedback on the final version of the draft.  We
would like to thank Nick Evans, for stimulating conversations 
which gave the start to the present work.  We
also thank M.Caldarelli, M.Hanada, G. Hartnett, K. Skenderis for
discussions.  Finally, we would like to thank an anonymous referee 
for pointing out several improvements in our narration. 
We acknowledge the use of the IRIDIS High Performance
Computing Facility, and its associated support services at the
University of Southampton.  FA acknowledges support from STFC through
Consolidated Grant ST/L000296/1.  FS received funding from the
European Research Council under the European Community Seventh
Framework Programme (FP7/2007-2013) ERC grant agreement No 279757.
\end{acknowledgments}

%%%%%%%%%%%%%%%%%%%%%%%%%%%%%%%%%%%%%%%%%%%%%%%%%%%%%%%
\section{Appendix}
%%%%%%%%%%%%%%%%%%%%%%%%%%%%%%%%%%%%%%%%%%%%%%%%%%%%%%%

%==================================================================================
\subsection{A.~ Determination of the quasi-normal frequencies}
%==================================================================================

For completeness, we describe in detail the algorithm we used to
determine, at fixed $h$ and $\overline{\epsilon}$, the values of the
parameters entering the ansatz
\beq\label{fit_appendix}
\OO(t) = \OO_{\infty} + e^{-t/\tau } \Big( d + u \cos(\omega t +
\phi)\Big)~,
\eeq
of the lowest quasi-normal mode. The parameters are obtained by
minimizing the $\chi^2$:
\beq
\chi^2= \sum_{t=t_{min}}^{t_{max}} \frac{ \OO(t)-
  \langle\OO\rangle(t)}{\sigma^2(t) }~.
\eeq
where $\langle\OO\rangle(t)$ is defined in~\eqref{def_OOt_num}, and
$\sigma(t)$ is the error on $\langle\OO\rangle(t)$ estimated via
jackknife analysis.

Setting the quench time at $t=0$ for convenience, the time $t_{\max}$
is chosen where the $\langle\OO\rangle$ has equilibrated.  In
practise, we fix it to the time after which the physical oscillations
of $\langle\OO\rangle(t)$ are indistinguishable from the statistical
fluctuations.  In this way we minimize the contribution of irrelevant
noise entering the $\chi^2$ at late times.
Moreover, in order to isolate the lowest quasi-normal mode we have to choose a
$t_{min}$ large enough such that the contributions to $\OO(t)$ coming from
faster decaying (higher) quasi-normal modes are negligible. 
This is determined by increasing $t_{min}$ progressively, from zero
towards $t_{max}$, until the estimated frequencies are stable against
a further variations of $t_{min}$. After subtracting the lowest
quasi-normal mode from the signal, we could in principle repeat the
procedure to isolate the next quasi-normal mode, and so on. A more
sophisticated approach would be actually needed to improve the stability
of the results. For simplicity, in this work we have focused mainly on
the lowest one.

The errors on the frequencies are estimated from the spread of
distribution of the fit parameters obtained from the different
jacknife samples used.

%==================================================================================
\subsection{B.~ On Quenches far-from-equilibrium}  
%==================================================================================

\begin{figure}[t]
\includegraphics[scale=0.38]{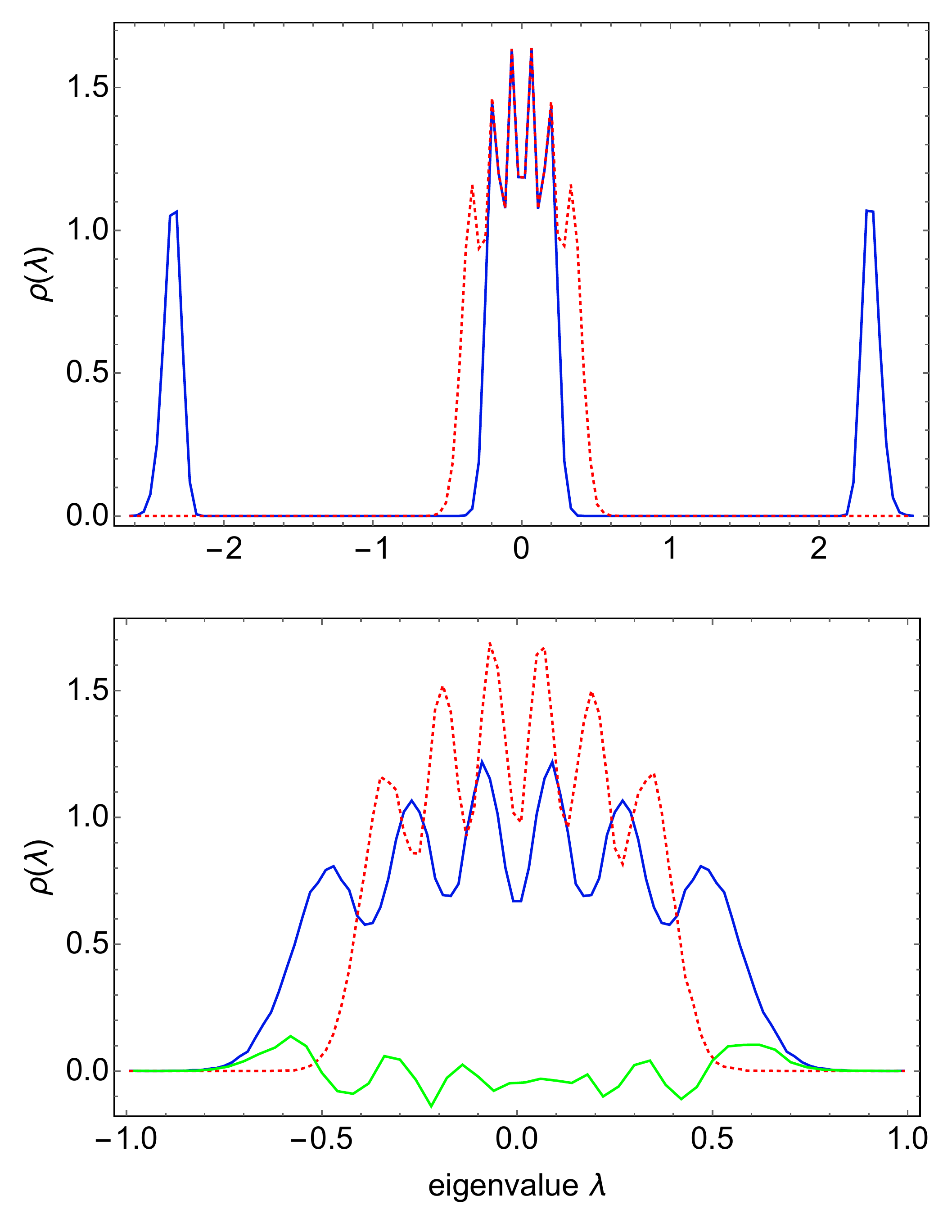}
\vskip 0.cm
\caption{The distributions $\rho^{eig}_{\mathfrak{p}=4}$ for $N=6$ at
  different instants of time, $h=0.3$.  Top panel: the dotted red
  curve corresponds to $t^{-}_{quench}$ and the solid blue curve to
  $t^{+}_{quench}$, with $\epsilon=2$.  Bottom panel: the solid curves
  correspond to the $\rho^{eig}_{4}$ (blue) and
  $\rho^{eig}_{4}-\rho^{eig}_5$ (green) at the new equilibrium in
  $\EE'$, the dotted red curve is taken from the top panel. }
\label{Fig7}
\end{figure}

In this section, we briefly describe edge-type quenches, in which
$\mathcal{E}'$ is a far-from-equilibrium configuration compared to
$\mathcal{E}$.  We also use these type of quenches to illustrate basic
consequences of the explicit symmetry breaking which persists in the
ensemble at the new equilibrium.

By construction, far-from-equilibrium configurations can be easily
engineered by increasing the strength of the quench parameters.
Considering a large value of the quench parameter $\epsilon$, the
distribution of eigenvalues of $\rho^{eig}_{\mathfrak{p}}$ at
$t^+_{quench}$ can be deformed as follows: $\rho^{eig}_{\mathfrak{p}}$
will contain three disconnected pieces, two outer peaks, whose support
is very well separated, and a central ``bubble'' of eigenvalues.
In the top panel of Figure~\ref{Fig7} we show a snapshot of the
distribution $\rho^{eig}_{\mathfrak{p}}$ at $t^{-}_{quench}$ and
$t^+_{quench}$ for $N=6$ in the case of $\epsilon=2$ and $h=0.3$.  It
is worth emphasizing that this particular profile of
$\rho^{eig}_{\mathfrak{p}}$ provides a different realization of the
framework of~\cite{Aoki:2015uha}.

The time evolution of the distribution after the quench proceeds as
the intuition suggests: The two outer peaks start moving towards the
central region until a new equilibrium is reached.  At the new
equilibrium $\rho^{eig}_{\mathfrak{p}}$ has a unique support.  The
apparent attractive force between the outer and the central part of the density of
eigenvalues should not be interpreted as originating from an
attractive force in the microscopic potential. In fact, the flow is
expansive. Instead, the nature of the force is statistical, and it appears in the chaotic regime as a 
consequence of the phase space getting populated  
according to the equilibrium distribution.

We can explore the
symmetry breaking pattern after the quench by analysing the equilibrium distributions of
$\rho^{eig}_{\mathfrak{p}}$ and $\rho^{eig}_{ I\neq \mathfrak{p}}$, as
shown in the bottom panel of Figure~\ref{Fig7}.  The green histogram
is the difference between $\rho^{eig}_{\mathfrak{p}}$ and
$\rho^{eig}_{\i\neq \mathfrak{p}}$, and it is non vanishing. The
symmetry breaking in the case of close-to-equilibrium quenches has
less prominent features, but can be seen more clearly upon increasing
the volume of the statistics.
Finally, since the change in the conserved charges is non perturbative
the expectation values of the charged operators $\mathcal{N}_{i=1,2}$
do not show a simple quasi-normal oscillation. Instead,
they display a more complicated behavior in which the under and
over-damping behaviors are equally mixed. A more sophisticated
analysis than the one used in Appendix {\bf A} would be needed to decompose
the signal into a sum of quasi-normal modes.

\vspace{0.5cm}

%%%%%%%%%%%%%%%%%%%


\begin{thebibliography}{9}

\vspace{0.5cm}

\bibitem{Dyson}
F.~J.~Dyson, J.\ Math.\ Phys.\ {\bf 3},\ 1191,\ {(1962)}.
  

%\cite{Blaizot:2009ex}
\bibitem{Blaizot:2009ex} 
  J.~P.~Blaizot and M.~A.~Nowak,
%  ``Universal shocks in random matrix theory''
  Phys. Rev. E {\bf 82}, 051115, (2010) 
  
  
 %\cite{Berenstein:2002jq}
\bibitem{Berenstein:2002jq} 
  D.~E.~Berenstein, J.~M.~Maldacena and H.~S.~Nastase,
%  ``Strings in flat space and pp waves from N=4 superYang-Mills,''
  JHEP {\bf 0204}, 013 (2002)
  doi:10.1088/1126-6708/2002/04/013
  [hep-th/0202021].
  %%CITATION = doi:10.1088/1126-6708/2002/04/013;%%
  %1560 citations counted in INSPIRE as of 08 Apr 2016
 
 
 %\cite{Kim:2003rza}
\bibitem{Kim:2003rza} 
  N.~Kim, T.~Klose and J.~Plefka,
%  ``Plane wave matrix theory from N=4 superYang-Mills on R x S**3,''
  Nucl.\ Phys.\ B {\bf 671}, 359 (2003)
  doi:10.1016/j.nuclphysb.2003.08.019
  [hep-th/0306054].
  %%CITATION = doi:10.1016/j.nuclphysb.2003.08.019;%%
  %81 citations counted in INSPIRE as of 08 Apr 2016

 
 %\cite{Gur-Ari:2015rcq}
\bibitem{Gur-Ari:2015rcq} 
  G.~Gur-Ari, M.~Hanada and S.~H.~Shenker,
%  ``Chaos in Classical D0-Brane Mechanics,''
  JHEP {\bf 1602}, 091 (2016)
  doi:10.1007/JHEP02(2016)091
  [arXiv:1512.00019 [hep-th]].
  %%CITATION = doi:10.1007/JHEP02(2016)091;%%
  %12 citations counted in INSPIRE as of 28 Jun 2016

 
 %\cite{Asplund:2011qj,Asplund:2012tg}
\bibitem{Asplund:2011qj} 
  C.~Asplund, D.~Berenstein and D.~Trancanelli,
%  ``Evidence for fast thermalization in the plane-wave matrix model,''
  Phys.\ Rev.\ Lett.\  {\bf 107}, 171602 (2011)
  doi:10.1103/PhysRevLett.107.171602
  [arXiv:1104.5469 [hep-th]].
  %%CITATION = doi:10.1103/PhysRevLett.107.171602;%%
  %44 citations counted in INSPIRE as of 08 Apr 2016


%\cite{Asplund:2012tg}
\bibitem{Asplund:2012tg} 
  C.~T.~Asplund, D.~Berenstein and E.~Dzienkowski,
%  ``Large N classical dynamics of holographic matrix models,''
  Phys.\ Rev.\ D {\bf 87}, no. 8, 084044 (2013)
  doi:10.1103/PhysRevD.87.084044
  [arXiv:1211.3425 [hep-th]].
  %%CITATION = doi:10.1103/PhysRevD.87.084044;%%
  %30 citations counted in INSPIRE as of 08 Apr 2016


%\cite{Gopakumar:1994iq}
\bibitem{Gopakumar:1994iq} 
  R.~Gopakumar and D.~J.~Gross,
  %``Mastering the master field,''
  Nucl.\ Phys.\ B {\bf 451}, 379 (1995)
  doi:10.1016/0550-3213(95)00340-X
  [hep-th/9411021].
  %%CITATION = doi:10.1016/0550-3213(95)00340-X;%%
  %107 citations counted in INSPIRE as of 16 Oct 2016
 
 
%\cite{Gomis:2008qa}
\bibitem{Gomis:2008qa} 
  J.~Gomis, S.~Matsuura, T.~Okuda and D.~Trancanelli,
  %``Wilson loop correlators at strong coupling: From matrices to bubbling geometries,''
  JHEP {\bf 0808}, 068 (2008)
  doi:10.1088/1126-6708/2008/08/068
  [arXiv:0807.3330 [hep-th]].
  %%CITATION = doi:10.1088/1126-6708/2008/08/068;%%
  %35 citations counted in INSPIRE as of 16 Oct 2016
 
 
 %\cite{Buchel:2013id}
\bibitem{Buchel:2013id} 
  A.~Buchel, J.~G.~Russo and K.~Zarembo,
  %``Rigorous Test of Non-conformal Holography: Wilson Loops in N=2* Theory,''
  JHEP {\bf 1303}, 062 (2013)
  doi:10.1007/JHEP03(2013)062
  [arXiv:1301.1597 [hep-th]].
  %%CITATION = doi:10.1007/JHEP03(2013)062;%%
  %38 citations counted in INSPIRE as of 16 Oct 2016
 
 
 %\cite{Benini:2015eyy}
\bibitem{Benini:2015eyy} 
  F.~Benini, K.~Hristov and A.~Zaffaroni,
  %``Black hole microstates in AdS$_{4}$ from supersymmetric localization,''
  JHEP {\bf 1605}, 054 (2016)
  doi:10.1007/JHEP05(2016)054
  [arXiv:1511.04085 [hep-th]].
  %%CITATION = doi:10.1007/JHEP05(2016)054;%%
  %16 citations counted in INSPIRE as of 16 Oct 2016
 
 
 %\cite{Bantilan:2012vu}
\bibitem{Bantilan:2012vu} 
  H.~Bantilan, F.~Pretorius and S.~S.~Gubser,
  %``Simulation of Asymptotically AdS5 Spacetimes with a Generalized Harmonic Evolution Scheme,''
  Phys.\ Rev.\ D {\bf 85}, 084038 (2012)
  doi:10.1103/PhysRevD.85.084038
  [arXiv:1201.2132 [hep-th]].
  %%CITATION = doi:10.1103/PhysRevD.85.084038;%%
  %70 citations counted in INSPIRE as of 17 Oct 2016
 
 
 %\cite{Costa:2014wya}
\bibitem{Costa:2014wya} 
  M.~S.~Costa, L.~Greenspan, J.~Penedones and J.~Santos,
  %``Thermodynamics of the BMN matrix model at strong coupling,''
  JHEP {\bf 1503}, 069 (2015)
  doi:10.1007/JHEP03(2015)069
  [arXiv:1411.5541 [hep-th]].
  %%CITATION = doi:10.1007/JHEP03(2015)069;%%
  %6 citations counted in INSPIRE as of 28 Jul 2017
 

%\cite{Banks:1996vh}
\bibitem{Banks:1996vh} 
  T.~Banks, W.~Fischler, S.~H.~Shenker and L.~Susskind,
%  ``M theory as a matrix model: A Conjecture,''
  Phys.\ Rev.\ D {\bf 55}, 5112 (1997)
  doi:10.1103/PhysRevD.55.5112
  [hep-th/9610043].
  %%CITATION = doi:10.1103/PhysRevD.55.5112;%%
  %2520 citations counted in INSPIRE as of 29 Jun 2016   
    
      
%\cite{Omelyan}
\bibitem{Omelyan} 
  I.~P.~Omelyan, I.~M.~Mryglod, and R.~Folk,
%  ``Optimized Verlet-like algorithms for molecular dynamics simulations,''
  Phys.\ Rev.\ E {\bf 65}, 056706 (2002)
  doi:10.1103/PhysRevE.65.056706
  [arXiv:cond-mat/0110438].
  
  
%  
\bibitem{NoteLyapu}
   Rainer Klages,
  ``Introduction to Dynamical Systems,''
   School of Mathematical Sciences Queen Mary, University of London.
 
 
%cite{TGUexact}   
\bibitem{TGUexact}
 K.~Ho, J.M.~Kahn, 
%``Statistics of Group Delays in Multimode Fiber with Strong Mode Coupling, Supplement,''
 Journal of Lightwave Technology, vol. 29, pp. 3119-3128, 2011 
 [arXiv:1104.4527v2 [physics.optics]]
  

 %\cite{Aoki:2015uha}
\bibitem{Aoki:2015uha} 
  S.~Aoki, M.~Hanada and N.~Iizuka,
%  ``Quantum Black Hole Formation in the BFSS Matrix Model,''
  JHEP {\bf 1507}, 029 (2015)
  doi:10.1007/JHEP07(2015)029
  [arXiv:1503.05562 [hep-th]].
  %%CITATION = doi:10.1007/JHEP07(2015)029;%%
  %3 citations counted in INSPIRE as of 21 Jul 2016


%\cite{Maldacena:1997re}
\bibitem{Maldacena:1997re} 
  J.~M.~Maldacena,
  %``The Large N limit of superconformal field theories and supergravity,''
  Int.\ J.\ Theor.\ Phys.\  {\bf 38}, 1113 (1999)
  [Adv.\ Theor.\ Math.\ Phys.\  {\bf 2}, 231 (1998)]
  doi:10.1023/A:1026654312961
  [hep-th/9711200].
  %%CITATION = doi:10.1023/A:1026654312961;%%
  %12041 citations counted in INSPIRE as of 23 Aug 2016


%\cite{D'Hoker:2002aw}
\bibitem{D'Hoker:2002aw} 
  E.~D'Hoker and D.~Z.~Freedman,
  %``Supersymmetric gauge theories and the AdS / CFT correspondence,''
  hep-th/0201253.
  %%CITATION = HEP-TH/0201253;%%
  %448 citations counted in INSPIRE as of 22 Sep 2016


%\cite{Witten:1998qj}
\bibitem{Witten:1998qj} 
  E.~Witten,
  %``Anti-de Sitter space and holography,''
  Adv.\ Theor.\ Math.\ Phys.\  {\bf 2}, 253 (1998)
  [hep-th/9802150].
  %%CITATION = HEP-TH/9802150;%%
  %7919 citations counted in INSPIRE as of 23 Aug 2016


%\cite{Bekenstein:1973ur} %\cite{Gibbons:1976ue}
\bibitem{Bekenstein:1973ur} 
  J.~D.~Bekenstein,
  %``Black holes and entropy,''
  Phys.\ Rev.\ D {\bf 7}, 2333 (1973).
  doi:10.1103/PhysRevD.7.2333
  %%CITATION = doi:10.1103/PhysRevD.7.2333;%%
  %3359 citations counted in INSPIRE as of 23 Aug 2016


%\cite{Gibbons:1976ue}
\bibitem{Gibbons:1976ue} 
  G.~W.~Gibbons and S.~W.~Hawking,
  %``Action Integrals and Partition Functions in Quantum Gravity,''
  Phys.\ Rev.\ D {\bf 15}, 2752 (1977).
  doi:10.1103/PhysRevD.15.2752
  %%CITATION = doi:10.1103/PhysRevD.15.2752;%%
  %1965 citations counted in INSPIRE as of 23 Aug 2016

\bibitem{Malda}
For a quick comparison with the notation of \cite{Maldacena:1997re}, 
our $M$ is such that $M=U_0^4/L^4\propto \mu\, g/N$. 


%\cite{Berti:2009kk}
\bibitem{Berti:2009kk} 
  E.~Berti, V.~Cardoso and A.~O.~Starinets,
  %``Quasinormal modes of black holes and black branes,''
  Class.\ Quant.\ Grav.\  {\bf 26}, 163001 (2009)
  doi:10.1088/0264-9381/26/16/163001
  [arXiv:0905.2975 [gr-qc]].
  %%CITATION = doi:10.1088/0264-9381/26/16/163001;%%
  %502 citations counted in INSPIRE as of 26 Aug 2016



%\cite{Liu:2003px}
\bibitem{Liu:2003px} 
%\cite{Liu:2004it}
%\bibitem{Liu:2004it} 
  J.~T.~Liu and W.~A.~Sabra,
  %``Mass in anti-de Sitter spaces,''
  Phys.\ Rev.\ D {\bf 72}, 064021 (2005)
  doi:10.1103/PhysRevD.72.064021
  [hep-th/0405171].
  %%CITATION = doi:10.1103/PhysRevD.72.064021;%%
  %41 citations counted in INSPIRE as of 16 Sep 2016


%\cite{Witten:1998zw}
\bibitem{Witten:1998zw} 
  E.~Witten,
  %``Anti-de Sitter space, thermal phase transition, and confinement in gauge theories,''
  Adv.\ Theor.\ Math.\ Phys.\  {\bf 2}, 505 (1998)
  [hep-th/9803131].
  %%CITATION = HEP-TH/9803131;%%
  %2415 citations counted in INSPIRE as of 24 Aug 2016


%\cite{Cvetic:2000nc}
\bibitem{Cvetic:2000nc} 
  M.~Cvetic, H.~Lu, C.~N.~Pope, A.~Sadrzadeh and T.~A.~Tran,
  %``Consistent SO(6) reduction of type IIB supergravity on S**5,''
  Nucl.\ Phys.\ B {\bf 586}, 275 (2000)
  doi:10.1016/S0550-3213(00)00372-2
  [hep-th/0003103].
  %%CITATION = doi:10.1016/S0550-3213(00)00372-2;%%
  %120 citations counted in INSPIRE as of 24 Aug 2016


%\cite{Skenderis:2006di}
\bibitem{Skenderis:2006di} 
  K.~Skenderis and M.~Taylor,
  %``Holographic Coulomb branch vevs,''
  JHEP {\bf 0608}, 001 (2006)
  doi:10.1088/1126-6708/2006/08/001
  [hep-th/0604169].
  %%CITATION = doi:10.1088/1126-6708/2006/08/001;%%
  %37 citations counted in INSPIRE as of 25 Aug 2016


%\cite{Bianchi:2001kw}
\bibitem{Bianchi:2001kw} 
  M.~Bianchi, D.~Z.~Freedman and K.~Skenderis,
  %``Holographic renormalization,''
  Nucl.\ Phys.\ B {\bf 631}, 159 (2002)
  doi:10.1016/S0550-3213(02)00179-7
  [hep-th/0112119].
  %%CITATION = doi:10.1016/S0550-3213(02)00179-7;%%
  %390 citations counted in INSPIRE as of 11 Mar 2017



%\cite{Horowitz:1999jd}
\bibitem{Horowitz:1999jd} 
  G.~T.~Horowitz and V.~E.~Hubeny,
  %``Quasinormal modes of AdS black holes and the approach to thermal equilibrium,''
  Phys.\ Rev.\ D {\bf 62}, 024027 (2000)
  doi:10.1103/PhysRevD.62.024027
  [hep-th/9909056].
  %%CITATION = doi:10.1103/PhysRevD.62.024027;%%
  %514 citations counted in INSPIRE as of 23 Aug 2016  
  
    
  %\cite{Berti:2004ju}
\bibitem{Berti:2004ju} 
  E.~Berti, V.~Cardoso and J.~P.~S.~Lemos,
%  ``Quasinormal modes and classical wave propagation in analogue black holes,''
  Phys.\ Rev.\ D {\bf 70}, 124006 (2004)
  doi:10.1103/PhysRevD.70.124006
  [gr-qc/0408099].
  %%CITATION = doi:10.1103/PhysRevD.70.124006;%%
  %93 citations counted in INSPIRE as of 29 Jun 2016  
  
  
    %\cite{Horowitz:1996nw}
\bibitem{Horowitz:1996nw} 
  G.~T.~Horowitz and J.~Polchinski,
%  ``A Correspondence principle for black holes and strings,''
  Phys.\ Rev.\ D {\bf 55}, 6189 (1997)
  doi:10.1103/PhysRevD.55.6189
  [hep-th/9612146].
  %%CITATION = doi:10.1103/PhysRevD.55.6189;%%
  %475 citations counted in INSPIRE as of 03 Aug 2016
  

       %\cite{Berkowitz:2016jlq}
\bibitem{Berkowitz:2016jlq} 
E.~Berkowitz, E.~Rinaldi, M.~Hanada, G.~Ishiki, S.~Shimasaki and P.~Vranas,
        %  ``Precision lattice test of the gauge/gravity duality at large-$N$,''
arXiv:1606.04951 [hep-lat].
        %%CITATION = ARXIV:1606.04951;%%

%\cite{Catterall:2014vka}
\bibitem{Catterall:2014vka} 
  S.~Catterall, D.~Schaich, P.~H.~Damgaard, T.~DeGrand and J.~Giedt,
  %``N=4 Supersymmetry on a Space-Time Lattice,''
  Phys.\ Rev.\ D {\bf 90}, no. 6, 065013 (2014)
  doi:10.1103/PhysRevD.90.065013
  [arXiv:1405.0644 [hep-lat]].
  %%CITATION = doi:10.1103/PhysRevD.90.065013;%%
  %19 citations counted in INSPIRE as of 01 Nov 2016



%\cite{Verlinde:2010hp}
\bibitem{Verlinde:2010hp} 
  E.~P.~Verlinde,
  %``On the Origin of Gravity and the Laws of Newton,''
  JHEP {\bf 1104}, 029 (2011)
  doi:10.1007/JHEP04(2011)029
  [arXiv:1001.0785 [hep-th]].
  %%CITATION = doi:10.1007/JHEP04(2011)029;%%
  %551 citations counted in INSPIRE as of 16 Oct 2016
  

\end{thebibliography}
\end{document}